\documentclass[aps, prd, amsfonts, amssymb, amsmath, preprint, titlepage, nofootinbib]{revtex4-2}

\usepackage[utf8]{inputenc}
\usepackage[english]{babel}
\usepackage[pdftex, pdftitle={Trivializing maps for the O(3) model in 2D}, pdfauthor={Christopher Chamness and Daniel Kovner and Kostas Orginos}, hypertexnames=false]{hyperref} %

\usepackage{mathtools}
\DeclareMathOperator{\Tr}{Tr}
\usepackage{xcolor}
\usepackage{graphicx}
\usepackage{placeins}
\usepackage[T1]{fontenc}

\newcommand\pder[2][]{\ensuremath{\frac{\partial#1}{\partial#2}}} 
\newcommand{\tran}{^{\mkern-1.5mu\mathsf{T}}}
\newcommand*{\s}[1]{s^{(#1)}}
\newcommand*{\e}[1]{^{(#1)}}

\usepackage{subfiles} %

\begin{document}
\title{Trivializing maps for the \texorpdfstring{$\mathrm{O}(3)$}{O(3)} model in 2D}
\newcommand*{\WM}{Physics Department, College of William and Mary, Williamsburg, Virginia 23187, USA}\affiliation{\WM}
\author{Christopher Chamness}
    \email{cdchamness@wm.edu}
    \affiliation{\WM}
\author{Daniel Kovner}
    \email{dkovner@wm.edu}
    \affiliation{\WM}
\author{Kostas Orginos}
    \email[Contact author: ]{kostas@wm.edu}
    \affiliation{\WM}
    
\date{\today}

\begin{abstract}
We construct trivializing maps for the two-dimensional nonlinear $\mathrm{O}(3)$ sigma model from gradient flow.
Trivializing maps, proposed to accelerate lattice Monte Carlo simulations, are changes of variables that trivialize the
path-integral measure. We compute the flow action perturbatively in the flow time through second order, organized by
eigenfunctions of the Lie Laplacian, and derive adjoint-sensitivity equations for transformed-HMC forces and parameter
gradients. Numerical experiments confirm the expected scaling of the flow-time truncations and show their limitations at
large volumes and near criticality. In exact transformed HMC the map substantially reduces autocorrelation times in
trajectory units at fixed trajectory length, but the cost of the transformed force and endpoint action prevents a
wall-clock speedup in the present implementation. Nonperturbative tuning of the flow-action coefficients reduces the action
mismatch by $12$--$28\%$ for $L\ge 6$ and extends the usable volume window at fixed $\xi/L$ by one step in $L$, but does
not overcome the exponential decay of the reweighting effective sample size with volume within the perturbative operator
basis. Surrogate effective actions, fitted by minimizing a Kullback--Leibler divergence or the variance of the action
mismatch, extend this window modestly but fail as multi-timestepping force surrogates, because these scalar objectives
impose no direct constraint on the force. The results provide a framework for trivializing flows in nonlinear sigma models
and identify the bottlenecks to be addressed before applying this approach to lattice QCD.
\end{abstract}

\maketitle

\section{Introduction}
\label{sec:intro}

Efficient sampling of lattice field theories remains a central challenge in computational physics. Standard algorithms such as Hybrid Monte Carlo (HMC) become increasingly inefficient near criticality because of growing autocorrelation times. One possible way forward, proposed by L\"uscher~\cite{Luscher:2009eq}, is to apply a trivializing map: a change of variables that transforms the path-integral measure into a trivial distribution. In this framework, the dynamics of the original fields can be reconstructed by inverting the transformation, while the trivialized variables can be sampled efficiently.

L\"uscher introduced the idea in the context of lattice gauge theory, using gradient flow to construct the map perturbatively in the flow time. For a systematic account of the Wilson, or gradient, flow and the properties used in Sec.~\ref{sec:O3model}, see Ref.~\cite{Luscher:2010iy}. While the method offered analytic control and a well-defined connection to the underlying theory, its practical impact has been limited. An application to the $CP(N-1)$ model~\cite{Engel:2011re} showed that the leading-order flow action gives only a small improvement, which is compensated by the additional computational overhead. Nonetheless, the framework remains compelling: trivializing maps offer a principled way to reduce critical slowing down, and their systematic construction is closely connected to renormalization group ideas. L\"uscher has also constructed trivializing maps order by order in the couplings, which yields an instantaneous form of stochastic perturbation theory for a wide class of lattice field theories, including lattice QCD~\cite{Luscher:2014mka}, demonstrating that the concept can be realized outside the strict gradient-flow setting. More recently, machine-learning methods, in particular normalizing flows, have been used to learn approximate trivializing maps, with promising results for scalar and gauge field theories~\cite{Kanwar:2020xzo,Albergo:2019eim,DelDebbio:2021qwf,Nicoli:2020njz,Boyda:2020hsi,Cheng:2025khv,Bauer:2024byr,Abbott:2025kvi,Gerdes:2022eve,Bacchio:2022vje}; learned field transformations have also been used as changes of variables inside HMC~\cite{Albandea:2023wgd,Jin:2022bgq}.

In this work, we study the construction of trivializing maps for the two-dimensional nonlinear $\mathrm{O}(3)$ sigma model, a simple theory with a compact target space and nontrivial dynamics. The $\mathrm{O}(3)$ model provides a useful testing ground: it shares structural features with non-Abelian gauge theories while remaining computationally simpler, and its path-integral measure can be represented either in terms of $S^2$ spins or $\mathrm{SO}(3)$ rotations. 
Although cluster algorithms, such as the Wolff algorithm~\cite{Wolff:1988uh}, eliminate critical slowing down in this model, we compare against HMC because it is the standard simulation algorithm for lattice QCD and many other lattice field theories.
We pursue both perturbative and nonperturbative approaches. Perturbatively, we expand the flow action in powers of flow time and explicitly compute contributions up to second order, identifying the role of eigenfunctions of the Lie Laplacian. Nonperturbatively, we formulate the determination of the coefficients as a variational problem, based on the Kullback--Leibler divergence or on the variance of the action mismatch, and test it together with an adjoint-sensitivity method for efficient gradient evaluation~\cite{Cao2003AdjointSA,Gerdes:2024rjk}. Finally, we present numerical tests that quantify the accuracy of truncated flows, their dependence on lattice size and coupling, and their performance inside exact transformed HMC.

This paper extends our preliminary report~\cite{Chamness:2024fob}, in which the gradient-flow trivializing map of the $\mathrm{O}(3)$ model in $\mathrm{SO}(3)$ variables and its flow action through $\mathcal{O}(t^2)$ were first presented. The main contributions of this work are fourfold. First, we give a complete formulation of gradient-flow trivializing maps for the $\mathrm{O}(3)$ sigma model in $\mathrm{SO}(3)$ variables. Second, we derive the perturbative flow action through second order in flow time and organize the result using eigenfunctions of the Lie Laplacian. Third, we derive continuum adjoint-sensitivity equations on $\mathrm{SO}(3)$ for transformed-HMC forces and for variational derivatives of flow-action parameters, extending the adjoint-state treatment of learned gradient flows in $\mathrm{SU}(N)$ gauge theories~\cite{Bacchio:2022vje,Gerdes:2024rjk}. Finally, we test the resulting maps numerically,  showing that they are approximately trivializing for a limited range of volumes and couplings $\beta$. When used to construct  a transformed HMC  algorithm,  autocorrelations are reduced by roughly an order of magnitude in trajectory units at a fixed trajectory length. However, the transformed HMC is not yet competitive with direct HMC in wall-clock time for the implementation studied here. Earlier studies reached similar conclusions for the leading-order map in the $CP(N-1)$ model~\cite{Engel:2011re}, for learned trivializing flows in two-dimensional $\phi^4$ theory, which reduced autocorrelation times but did not improve the scaling of the cost toward the continuum relative to HMC~\cite{Albandea:2023wgd}, and for maps constructed from Schwinger--Dyson equations in $\mathrm{SU}(3)$ gauge theory, where the algorithmic overhead outweighed the faster decorrelation~\cite{Boyle:2022xor}.

\section{The nonlinear \texorpdfstring{$\mathrm{O}(3)$}{O(3)} model}
\label{sec:O3model}

We begin by reviewing the lattice formulation of the two-dimensional $\mathrm{O}(3)$ 
nonlinear sigma model and its associated integration measure. This will provide 
the foundation for the construction of trivializing maps in the following sections. 

The model on a square lattice is defined by the action
   \begin{align}
        {\cal S}[s] = -\beta \sum_{x,\hat\mu>0} s_x\tran s_{x+\hat\mu} \,,
   \end{align}
   with $s_x \in S^2$.
This action can equivalently be written as
 \begin{align}
        {\cal S}[s] = -\frac{1}{2}\beta \sum_{x} s_x\tran s^{(1)}_x 
   \end{align}
   where
   \begin{align}
       s^{(1)}_x = \sum_\mu s_{x+\hat\mu}
   \end{align}
   and the sum over $\mu$ runs over both positive and negative directions. This notation will be useful below.
   
   The corresponding probability distribution is
   \begin{align}
       p[s] = \frac{ e^{-{\cal S}[s]}}{\cal Z},
   \end{align}
   with partition function
   \begin{align}
     {\cal Z} = \int {\cal D} s \, e^{-{\cal S} [s] }\,.
   \end{align}
   The integration measure can be defined over unit-norm $\mathbb{R}^3$ vectors, i.e., the compact manifold $S^2$, as
   \begin{align}
       \mathcal{D} s \equiv \prod_x \left[d^3r_x\,\delta(r_x\tran r_x -1) \right]
   \end{align}
  where $r_x \in \mathbb{R}^3$. While this definition is standard, the presence of the 
constraint can complicate some calculations. An equivalent and often more convenient 
formulation uses $\mathrm{SO}(3)$ variables. This formulation is better suited for our application, so we describe it in more detail.

  Consider a general integral over a single spin of the form
  \begin{align}
      I(h\tran h) = \int ds f(h\tran s)
  \end{align}
  where $h$ is an arbitrary vector in $\mathbb{R}^3$. Up to an overall normalization given by the volume of the stabilizer subgroup $\mathrm{SO}(2)$, the same integral can be written as
  \begin{align}
      I(h\tran h)\propto \int_{\mathrm{SO}(3)} d U f(h\tran U s_0)
  \end{align}
  with $U\in \mathrm{SO}(3)$ and $s_0$ an arbitrary fixed element of $S^2$. Accordingly, the probability distribution and partition function of the $\mathrm{O}(3)$ model 
can be expressed as
\begin{align}
    p[s]\,{\cal D}s =& \frac{1}{\cal Z'} e^{-{\cal S }[U\cdot s_0]}{\cal D} U\\
    {\cal Z'} =&  \int_{\mathrm{SO}(3)} {\cal D} U\, e^{-{\cal S }[U\cdot s_0]}
\end{align}
where $\mathcal{D}U$ denotes the product of Haar measures of $\mathrm{SO}(3)$ over 
all lattice sites. Since correlation functions computed with this measure are identical 
to those obtained from the $S^2$ definition, we may freely use the notation
\begin{align}
     p[s] {\cal D} s =& \frac{1}{\cal Z'} e^{-{\cal S }[U\cdot s_0]}{\cal D} U = \frac{1}{\cal Z} e^{-{\cal S }[s]}{\cal D} s
\end{align}
with the understanding that $s_x = U_x \cdot s_0$. 

Although all expressions can be written in terms of $S^2$ variables, the 
$\mathrm{SO}(3)$ formulation proves advantageous when discussing changes of variables 
and their Jacobians, which are central to the construction of trivializing maps.

\subsection{Hybrid Monte Carlo in \texorpdfstring{$\mathrm{SO}(3)$}{SO(3)} variables}
\label{sec:HMC}

The Hybrid Monte Carlo algorithm~\cite{Duane:1987de} introduces auxiliary momentum 
variables for each degree of freedom in a way that preserves the marginal probability 
distribution of the original fields. The momenta are taken to be Gaussian, so that 
integrating them out recovers the Boltzmann weight for the spins. The extended action 
is identified with a Hamiltonian, and the fields evolve along continuous trajectories 
that preserve it. An accept/reject step at the end of each trajectory compensates for 
integration errors and ensures exactness.

   We introduce a particular choice of Hamiltonian dynamics that is simple to 
implement and naturally preserves the unit-norm constraint on the $\mathrm{O}(3)$ spins. 
In analogy with our measure definition, the time evolution of a spin is described by an $\mathrm{SO}(3)$ rotation $R$ as

   \begin{align}
      s_{x,t} = R_{x,t} s_{x,0}
      \label{eq:O3rotation}
   \end{align}
   We now devise a Hamiltonian dynamics for the rotation matrices $R_{x,t}$ that belong to the $\mathrm{SO}(3)$ group. 
   As in $\mathrm{SU}(N)$ Yang--Mills theory, we introduce momenta that are elements of the Lie algebra $\mathfrak{so}(3)$ of $\mathrm{SO}(3)$.
   In other words, the momenta are traceless antisymmetric matrices that can be written in terms of the generators $L_a$ of $\mathrm{SO}(3)$ as
   \begin{align}
       p_x = \sum_{a=1}^3 L_a p^a_x
   \end{align}
   The Hamiltonian for the dynamics that governs the evolution of the fields $s$ is
\begin{align}
       H[p,s] =&  \frac{1}{4}\sum_x \Tr(p\tran_x p_x) 
       -\frac{1}{2}\beta \sum_{x}\left(s_x\tran s^{(1)}_x
       \right)\nonumber \\ \,
               =&  \frac{1}{2}\sum_{x,a} p^a_x p^a_x
       - \frac{1}{2}\beta \sum_{x}\left(s_x\tran s^{(1)}_x
       \right)
\end{align}
   Our conventions for the $\mathrm{SO}(3)$ generators are given in the Appendix. %
To derive the Hamiltonian dynamics, we introduce derivatives with respect to the vector field $s$.
  
Since we want to constrain the dynamics according to Eq.~\eqref{eq:O3rotation}, we define the Lie derivative as
\begin{align}
       \partial^a_x F[s] = \left.\frac{d}{dt} F[s_t]\right|_{t=0}\;\;,\;\;
       s_{y,t} =\left\{\begin{array}{lr}
            &  e^{L_a t} s_x\;,\; y=x\\
            & s_y \;,\; y\ne x
       \end{array}\right.\,,
\end{align}
where $F[s]$ is some functional of $s$, and \(s_t\) denotes the full field configuration, with site component
\((s_t)_y\equiv s_{y,t}\).
With this definition, we compute the gradient with respect to the $s$ fields and derive the Hamiltonian equations of motion:
\begin{align}
\dot{p}^a_x &= -\partial^a_x H \\
\dot{s}_x &= \sum_a (\partial_{p^a_x} H ) L_a s_x
\label{eq:Hamiltonian_z_ev}
\end{align}
   This form arises because the canonical pair in this Hamiltonian dynamics is ($p^a$,$\theta^a$), where $\theta^a$ defines the rotation
   \begin{align}
       R_{x,t} = e^{\sum_a \theta^a_{x,t}L_a}
   \end{align}
   that evolves $s_x$ in Hamiltonian evolution time. The coordinates $\theta^a_x$ are centred on the configuration at which the equations of motion are evaluated, $\theta_{x,t}=0$; any configuration along the trajectory can serve as $s_{x,0}$ in Eq.~\eqref{eq:O3rotation}.
   From the Hamiltonian equations for the canonical pair of coordinates, we obtain
   \begin{align}
   \dot \theta^a_x=\partial_{p^a_x} H = p^a_x
   \end{align}
   from which Eq.~\eqref{eq:Hamiltonian_z_ev} follows. At $\theta\neq0$, $\dot\theta^a_x$ and $p^a_x$ differ at first order in $\theta$; the form valid along the whole trajectory is $\dot R_{x,t}=p_xR_{x,t}$.
   
   The explicit form of the Hamiltonian equations is
   \begin{align}
       \dot p^a_x =& \beta\, {s^{(1)}_x}\tran L_a s_x\,,\\
       \dot p_x \equiv \sum_a L_a \dot p^a_x =& \beta\left(  s^{(1)}_x s_x\tran -
        s_x {s^{(1)}_x}\tran \right)\label{eq:Hamiltonian_explicit}\\
       \dot s_x =& \sum_a p^a_x L_a s_x = p_x s_x\,,
   \end{align}
   where $p_x \equiv \sum_a L_a p^a_x$.
   In deriving Eq.~\eqref{eq:Hamiltonian_explicit}, we used Eq.~\eqref{eq:compSO3}.
 
We have implemented this algorithm in Python and have reproduced the second-moment correlation length $\xi_2$ reported by Bietenholz {\em et al.}~\cite{Bietenholz:2018agd} at all eight $(L,\beta)$ values of their Table~1, as a check of correctness.

\subsection{Gradient flow}
\label{sec:flow}

Gradient flow provides a natural way to construct invertible changes of variables 
in the path integral. Given a functional $S_f[s]$, invariant under global 
$\mathrm{SO}(3)$ rotations, we define the flow equation
\begin{align}
  \dot s_{x,t} = \sum_a L_a \, \partial_x^a S_f[s] \, s_{x,t}\,,
  \label{eq:flow_equation_general}
\end{align}
with initial condition $s_{x,0} = \sigma_x$. Because the generator is 
antisymmetric, the flow preserves the norm, $s_{x,t}^\top s_{x,t} = 1$. 
The solution at flow time $t$ defines a map $s = F_t[\sigma]$. For this map to be potentially trivializing, the flow action must respect the global symmetries of the target action.  Invariance under global $\mathrm{O}(3)$ transformations requires that the flow action depend on the fields only through the inner products $s_x\tran s_y$; invariance under global $\mathrm{SO}(3)$ rotations alone would also allow triple products $s_x\tran(s_y\times s_z)$.  Rotational and translational invariance further constrain the form of the flow action. As a concrete example, let us consider the flows generated by the action itself.
Let 
\begin{align}
    S_{0f}[s] = -\frac{1}{2} \gamma_0 \sum_{x}\left[ s_x\tran s^{(1)}_x\right]\,,
\end{align}                                     where $\gamma_0$ is a coefficient that may depend on flow time. 
The corresponding flow is Eq.~\eqref{eq:flow_equation_general} with $S_f=S_{0f}$.  It is defined by considering infinitesimal $\mathrm{SO}(3)$ rotations. With this sign convention, forward integration in flow time maximizes the flow action. 

Before writing the generator explicitly, we introduce a new notation for the gradient with respect to $s$ as
\begin{align}
    {\bf \nabla}^s_x  \equiv \sum_a L_a \partial_x^a
\end{align}
With this notation, we have
\begin{align}
    {\bf \nabla}^s_x S_{0f} = -\gamma_0
    \left[  s^{(1)}_x s_x\tran -
        s_x {s^{(1)}_x}\tran \right]
\end{align}
Thus, the generator of the flow is
\begin{align}
    Z_x =  {\bf \nabla}^s_x S_{0f} = -\gamma_0
    \left[  s^{(1)}_x  s_x\tran -
        s_x {s^{(1)}_x}\tran \right] \;\;\; \mbox{\rm or}\;\;\;
    Z^a_x = \partial_x^a S_{0f}\,. 
\end{align}
The flow defined above preserves the norm of $s$ and, for either sign of $\gamma_0$, increases the flow action monotonically: at fixed $\gamma_0$, $dS_{0f}/dt=\sum_{x,a}(\partial^a_x S_{0f})^2\ge 0$, so the flow is a gradient ascent of $S_{0f}$. (The leading-order trivializing flow action of Sec.~\ref{sec:trivialFlow} is $S_{0f}$ with $\gamma_0=-\beta/4$.) By changing the sign of the generator, we obtain a flow in the opposite direction, which decreases the flow action monotonically. The flow equations can be integrated with standard Runge--Kutta methods suitably generalized by Munthe-Kaas to handle Lie group flows~\cite{MUNTHEKAAS1999115}. For efficient low-storage, commutator-free Lie-group schemes in flow integration, see also~\cite{Bazavov:2021pik}.

\subsection{Gradient flow maps and change of variables}
\label{sec:MAPS}

At each lattice site $x$, the field $s$ takes values $s_x$ on the sphere $S^{2}$. Maps from $S^2$ to $S^2$ can be described locally as rotations, i.e., multiplications by elements of $\mathrm{SO}(3)$, which could be a function of $s$. In other words, for a map $S^2 \rightarrow S^2$ where $s_x \rightarrow \sigma_x$, there exists $R_x \in \mathrm{SO}(3)$ so that
 \begin{align}
    \sigma_x = R_x\, s_x \,,
 \end{align}
 where $R_x$, which may depend on $s_x$, is an element of $\mathrm{SO}(3)$ defined only up to right multiplication by the $\mathrm{SO}(2)$ subgroup that leaves $s_x$ invariant, i.e., it represents a coset in $\mathrm{SO}(3)/\mathrm{SO}(2)$. This follows because the quotient $\mathrm{SO}(3)/\mathrm{SO}(2)$ is diffeomorphic to $S^{2}$.
 This fact allowed us to rewrite the integration measure over $S^2$ in terms of the $\mathrm{SO}(3)$ Haar measure.
 
 To change variables, we need one-to-one maps, which we define using a gradient flow:
 \begin{align}
    \dot s_x = \sum_a L_a \partial_x^a S_f[s] s_x\,.
    \label{eq:flow}
\end{align}
 Let $\sigma$ be the initial condition of the flow equation at flow time $t=0$. With this initial condition, the solution at flow time $t$ is unique and defines a one-to-one and invertible map ${\cal F}$: $[S^{2}]^{|\Lambda|} \rightarrow [S^{2}]^{|\Lambda|}$, where $|\Lambda|=L^2$ is the number of sites of the $L\times L$ lattice $\Lambda$. 
\begin{equation}
    s = {\cal F}[\sigma]
\end{equation}
Using the fixed vector $s_0$, we write $\sigma_x = V_x s_0$ with $V_x \in \mathrm{SO}(3)$ and
\begin{equation}
s = U \cdot s_0 = {\cal F}[\sigma] = {\cal F}[V s_0]
\end{equation}
where both $V_x$ and $U_x$ are in $\mathrm{SO}(3)$. This defines a map $\mathrm{SO}(3)\rightarrow\mathrm{SO}(3)$ that is one-to-one and invertible. The corresponding flow equation is
\begin{equation}
    \dot U_x = \sum_a L_a \partial_x^a S_f[U\cdot s_0] U_x = \sum_a L_a \partial_x^a S_f[s] U_x \,,
    \label{eq:Uflow}
\end{equation}
and initial condition $U_0 = V$. The generators of both the $S^2$ flow and the associated $\mathrm{SO}(3)$ flow are
\begin{align}
    Z_{x,t} = \sum_a L_a \partial_x^a S_f[s] \;\; {\rm or }\;\; Z^a_{x,t} = \partial_x^a S_f[s]
\end{align}
The solution of the flow equation with initial condition $V$ defines the associated map
\begin{align}
    U = {\cal G}[V]
\end{align}
Using this map, we define a change of variables from $U$  to $V$ (implying a change from $s$ to $\sigma$) resulting in a change of integration measure as
\begin{align}
    {\cal D} U = {\cal D} V \det({\cal J}[V]) \;\; {\rm  and}\;\; {\cal D} s = {\cal D} \sigma \det({\cal J}[V]) 
\end{align}
where $\cal J$ is the Jacobian of the transformation.  As shown in the Appendix, the Jacobian matrix is
\begin{align}
{\cal J}^{a,b}[V](x,y) = - \frac{1}{2}\Tr [\hat \partial_y^b U_x U\tran_x L_a]
\label{eq:JacobianMatrix}
\end{align}
where $\hat \partial_y^b$ are Lie derivatives with respect to the initial conditions of the flow $V$. Using the flow equations, one can show (see Appendix) that 
\begin{align}
      \frac{d \ln\det [{\cal J} ]}{dt} =&\; \sum_{x,a}[{\partial_x^a}]^2 S_f[s]
     \label{eq:jacobianODE}
\end{align}

 Integrating Eq.~\eqref{eq:jacobianODE} together with the flow equation gives the Jacobian determinant:
 \begin{align}
    \ln\det [{\cal J}_t ] = \int_0^t dt' \,\, \sum_{x,a} [\partial_x^a]^2 S_f[s_{t'},t']
 \end{align}
 which can be regarded either as a function of $\sigma$ or as a function of $s_t$. A useful observation is that the flow action $S_{0f}$ is an eigenfunction of the Lie Laplacian $\sum_{x,a} [\partial_x^a]^2$, so the computation of the Jacobian determinant simplifies further.
 \begin{align}
    \sum_{x,a} [\partial_x^a]^2 S_{0f}[s_{t}] =& - \gamma_0 \sum_{x,a}\partial_x^a\left[ {s^{(1)}_x}\tran L_a s_x\right]
    = -  \gamma_0 \sum_{x,a}\left[ {s^{(1)}_x}\tran L_a^2 s_x\right]\nonumber\\
     =&  2 \gamma_0 \sum_{x}\left[ {s^{(1)}_x}\tran  s_x\right] = - 4 S_{0f}[s_{t}]
 \end{align}
 where we used the $\mathrm{SO}(3)$ quadratic Casimir $C_F=2$.
 This result leads to the following Jacobian determinant for the $S_{0f}$ flow action:
 \begin{align}
     \ln\det [{\cal J}(U_t) ] = - 4 \int_0^t dt' \,\,  S_{0f}(s_{t'})
 \end{align}
 This is analogous to the $\mathrm{SU}(3)$ gauge-theory result of Ref.~\cite{Luscher:2009eq} using the Wilson action as a flow action and is a direct consequence of the fact that the chosen flow action is an eigenfunction of the Lie Laplacian.

\subsection{Trivializing flow}
\label{sec:trivialFlow}

Using the results from the previous section, we rewrite the effective action for the new variables $\sigma$ as
\begin{align}
    {\cal S}_{0}[\sigma,t] =& {\cal S}[{\cal F}_t[\sigma]]- \ln\det {\cal J}_t\Longleftrightarrow \nonumber \\
     {\cal S}_{0}[\sigma,t] =& {\cal S}[{\cal F}_t[\sigma]]-  \int_0^t dt' \,\, \sum_{x,a} [\partial_x^a]^2 S_f[s_{t'},t']
     \label{eq:exact_flowed_action}
\end{align}
As in Ref.~\cite{Luscher:2009eq}, the effective action ${\cal S}_{0}[\sigma,t]$ can be chosen to interpolate between the target action at $t=0$ and a trivial action at $t=1$ as follows:
\begin{align}
    {\cal S}_{0}[\sigma,t] = (1- t) {\cal S}[{\cal F}_t[\sigma]] + C(t)\,,
\end{align}
where $C(t)$ is a function independent of the fields satisfying the initial condition $C(0) = 0$. This is the simplest continuous interpolation between the target and trivial actions; however, other choices may be preferable for numerical efficiency. 
With this substitution, we obtain the equation
  \begin{align}
   - \int_0^t dt' \,\, \sum_{x,a} [\partial_x^a]^2 S_f[s_{t'},t'] + t {\cal S}[{\cal F}_t[\sigma]]  = C(t)
\end{align}
By differentiating with respect to flow time, we obtain
\begin{align}
 - \sum_{x,a} [\partial_x^a]^2 S_f[s_{t},t] + t \sum_{x,a} \partial_x^a {\cal S}[s_t] \partial_x^a S_f[s_t] = -  {\cal S}[s_t] + \dot C(t)\,,
 \label{eq:Luscher}
\end{align}
where $s_t = {\cal F}_t(\sigma)$, and the flow equations were used.  

A more general form of this equation can be derived if the time dependence of the effective action of the $\sigma$ variables remains unspecified. 
\begin{align}
 - \sum_{x,a} [\partial_x^a]^2 S_f[s_{t},t] +  \sum_{x,a} \partial_x^a {\cal S}[s_t] \partial_x^a S_f[s_t] =  \partial_t {\cal S}_0[\sigma,t]\,,
 \label{eq:EffActionPDE}
\end{align}
This equation can be viewed in two ways. Given the time-dependent effective action, the flow action that results in this effective action can be determined. However, if the flow action is given, this equation determines the time dependence and the functional form of the effective action. 

In this section we take the first point of view and determine the flow action for a given time dependence of the effective action. By introducing
the linear elliptic operator
\begin{align}
{\cal L}_t(\cdot) = -e^{t{\cal S}(s)}\sum_{x,a}\partial_x^a\left[ e^{-t{\cal S}(s)}\partial_x^a(\cdot)\right]\,
\end{align}
Eq.~\eqref{eq:Luscher} becomes
\begin{align}
   {\cal L}_t(S_f) = - {\cal S} + \dot C(t) 
\end{align}
The operator ${\cal L}_t$ is of Sturm--Liouville type and self-adjoint under the inner product
\begin{align}
    \langle G, F\rangle = \int {\cal D}s\,  e^{-t{\cal S}[s]}\, G[s] F[s]
\end{align}
As discussed in~\cite{Luscher:2009eq}, together with the elliptic property, this implies that the spectrum of ${\cal L}_t$ is discrete, has no accumulation points, and has differentiable, complete eigenfunctions. Furthermore, the only zero mode of ${\cal L}_t$ is the constant function $G_0[s] = 1/\langle1,1\rangle^{1/2}$. In addition, because Eq.~\eqref{eq:Luscher} is of Sturm--Liouville type, it has a solution, unique up to an additive constant, provided the right-hand side has no overlap with the zero mode of ${\cal L}_t$. Using the freedom in the constant $C(t)$, the zero mode can always be eliminated from the right-hand side and thus the class of gradient-flow maps contains the desired trivializing map. Furthermore, requiring the right-hand side to have no overlap with the zero mode results in
\begin{align}
    \dot{C}(t) = \frac{1}{\langle 1, 1\rangle } \int \mathcal{D}s\, {\cal S}[s]  e^{-t{\cal S}[s]}
\end{align}
with 
\begin{align}
    \langle 1, 1\rangle  = \int \mathcal{D}s\, e^{-t{\cal S}[s]} = \mathcal{Z}(t) \,.
\end{align}
For the $\mathrm{O}(3)$ action, $\mathcal{Z}(t)$ is just the partition function where the coupling is scaled by the flow time $t$. 
Furthermore, one obtains
\begin{align}
     {C}(t)  = \ln \left(\frac{\mathcal{Z}(0)}{\mathcal{Z}(t)}\right)
\end{align}
One way to solve Eq.~\eqref{eq:Luscher} is by expanding in a Taylor series at $t=0$~\cite{Luscher:2009eq}. Although such a solution has little chance of being accurate enough without going to very high order in the expansion, it is instructive to construct it and study its properties. 
The small flow-time expansion  of the flow action reads as
\begin{align}
    S_f[s] = S_f^{(0)}[s]  + \sum_{k=1}^\infty t^k S_f^{(k)}[s]
\end{align}
The unknown function $\dot C(t)$ can also be determined as a power series in $t$ as
\begin{align}
    \dot C(t) = C^{(0)} + \sum_{k=1}^{\infty} C^{(k)} t^k 
\end{align}
With this definition, the coefficients $C\e{k}$ are $(k+1)$ times the coefficients of order $k+1$ in the strong-coupling expansion of $-\ln \mathcal{Z}$ for the $\mathrm{O}(3)$ model, because they are defined through $\dot C(t)$. 
We solve for the flow action and the unknown constant $C(t)$ order by order in the flow-time expansion. Inserting both series into Eq.~\eqref{eq:Luscher} and equating the coefficients of $t^n$ for $n\ge1$ gives
\begin{align}
    - \sum_{a,x} [\partial^a_x]^2 S\e{n}_f = C\e{n} - \sum_{x,a}\partial_x^a {\cal S}\,\partial_x^a S\e{n-1}_f\,,
    \label{eq:FlowRecursion}
\end{align}
so that each order follows from the previous one by inverting the Lie Laplacian, with $C\e{n}$ chosen to remove the zero mode of the right-hand side. At zeroth order, we obtain
\begin{align}
    - \sum_{a,x} [\partial^a_x]^2 S\e{0}_f = -{\cal S} \Longleftrightarrow
    - \sum_{a,x} [\partial^a_x]^2 S\e{0}_f  = \frac{\beta}{2} \Psi\e{0}
\end{align}

where we introduced
\begin{align}
    \Psi\e{0} = \sum_x s_x\tran \s{1}_x \\
    \mathcal{S}[s] = -\frac{\beta}{2}\Psi\e{0}[s]
\end{align}
In addition, we have shown that the nearest-neighbor action is an eigenfunction of the Lie Laplacian; therefore $\Psi\e{0}$ is also an eigenfunction of the Lie Laplacian
\begin{align}
    -\sum_{a,x} [\partial^a_x]^2 \Psi\e{0} = 2 C_F \Psi\e{0}\,,
\end{align}
where $C_F=2$ is the eigenvalue of the quadratic Casimir of $\mathrm{SO}(3)$, as noted in Eq.~\eqref{eq:Casimir}.
Therefore the zeroth-order flow action is proportional to $\Psi\e{0}$
\begin{align}
    S\e{0}_f[s] = \frac{\beta}{4 C_F} \Psi\e{0}[s] \label{eq:ZerothOrder}\\
    C\e{0} = 0 \nonumber
\end{align}
The equation for the first-order flow action reads
\begin{align}
    - \sum_{a,x} [\partial^a_x]^2  S\e{1}_f   = C\e{1} - \sum_{x,a}\partial_x^a {\cal S}\partial_x^a S\e{0}_f
\end{align}
The constant $C\e{1}$ can be adjusted to remove the zero mode of the Lie Laplacian from the right-hand side, thereby allowing the computation of $S\e{1}_f$ by inverting the Lie Laplacian. A brief calculation leads to 
\begin{align}
    \sum_{x,a}\partial_x^a {\cal S}\partial_x^a S\e{0}_f =&
    - \frac{\beta^2}{2 C_F} \sum_{x} \left\{ [\s{1}_x]\tran \s{1}_x - 
   (s\tran_x \s{1}_x)(s\tran_x \s{1}_x)
    \right\}
\end{align}
which can be expanded in eigenfunctions of the Lie Laplacian. The required eigenfunctions are
\begin{align}
\Psi\e{2} =& \sum_x s\tran_x \s{2}_x \\
  - \sum_{a,x} [\partial^a_x]^2  \Psi\e{2} =& 2 C_F \Psi\e{2} 
\end{align}
where
\begin{align}
    \s{2}_x = \sum_\mu ( \s{1}_{x+\hat\mu} - s_x) \,,
\end{align}
and
\begin{align}
    \Psi\e{1,1}_l =& \sum_{x,\mu} (s\tran_x s_{x+\hat\mu})(s\tran_x  s_{x+\hat\mu}) - \frac{4|\Lambda|}{1+C_F} \\
     -  \sum_{a,x} [\partial^a_x]^2   \Psi\e{1,1}_l =& 4 (1+C_F) \Psi\e{1,1}_l \\
     \Psi\e{1,1} =& \sum_y (s\tran_y \s{1}_y)( s\tran_y\s{1}_y) -   \Psi\e{1,1}_l
    -\frac{1}{1+ C_F} \Psi\e{2} - 
    4|\Lambda|\frac{1}{1+C_F}\\
     - \sum_{a,x} [\partial^a_x]^2  \Psi\e{1,1} =&  (2+4 C_F) \Psi\e{1,1}\,.
\end{align}
With these eigenfunctions at hand, we have
\begin{align}
    \sum_{x,a}\partial_x^a {\cal S}\partial_x^a S\e{0}_f =& - \frac{\beta^2}{2 C_F} \left[ \frac{C_F}{1+ C_F} \Psi\e{2} 
    - \Psi\e{1,1}  -   \Psi\e{1,1}_l
     + 4|\Lambda|\frac{C_F}{1+C_F}
    \right] \,,
    \label{eq:S0contrFlow}
\end{align}
and
\begin{align}
     - \sum_{a,x} [\partial^a_x]^2  S\e{1}_f =  
     \frac{\beta^2}{2 C_F} \left[ \frac{C_F}{1+ C_F} \Psi\e{2} 
    - \Psi\e{1,1}  -   \Psi\e{1,1}_l
     + 
    4|\Lambda|\frac{C_F}{1+C_F}
    \right]
    + C\e{1}
\end{align}
which gives the following solution for the first-order flow action after setting $C_F=2$ 
\begin{align}
    S_f\e{1} =&  \frac{\beta^2}{8}\left[\frac{1}{3}\Psi\e{2} -
     \frac{1}{5} \Psi\e{1,1}   - 
      \frac{1}{6}\Psi\e{1,1}_l \right]\nonumber\\
     C\e{1}=& -\frac{2}{3}|\Lambda|\beta^2
\end{align}
Combining this with the leading-order term and dropping the irrelevant constant, we obtain
\begin{align}
    S_f[s] = \frac{\beta}{8} \Psi\e{0}  + \frac{\beta^2}{8}\left[\frac{1}{3}\Psi\e{2} -
     \frac{1}{5} \Psi\e{1,1}   - 
      \frac{1}{6}\Psi\e{1,1}_l \right]\,t\, + {\cal O}(t^2)\nonumber
\end{align}
By introducing the function
\begin{align}
   \tilde\Psi\e{1,1} = \sum_y  (s\tran_y \s{1}_y)(s\tran_y \s{1}_y) 
\end{align}
 the flow action reduces, up to a constant, to
\begin{align}
    S_f = \frac{\beta}{8} \Psi\e{0}  + \frac{\beta^2}{40}\left[2 \Psi\e{2} -
      \tilde\Psi\e{1,1}   + 
      \frac{1}{6}\Psi\e{1,1}_l \right]\,t\, + {\cal O}(t^2)\nonumber
\end{align}
which is simpler to implement numerically.
Note that $\tilde\Psi\e{1,1} $ is not an eigenfunction of the Lie Laplacian.

Using a computer-algebra program written in Python (with Rust components),\footnote{The program is publicly available at \url{https://github.com/cdchamness/O3OrderExpansion}.} we obtained the ${\cal O}(t^2)$ flow action, which was first reported in Ref.~\cite{Chamness:2024fob}. Both the ${\cal O}(t)$ and ${\cal O}(t^2)$ terms were independently verified by explicit hand calculation. The resulting flow action is

\begin{align}
    S_f =& \frac{\beta}{8}\left[1  - \frac{1489}{3000} \beta^2 t^2\right] \Psi\e{0}  + \frac{\beta}{40}\left[2 \Psi\e{2} -
      \tilde\Psi\e{1,1}   + 
      \frac{1}{6}\Psi\e{1,1}_l \right]\,\beta t \nonumber \\
      &+ \frac{\beta}{80}\bigg[\frac{29}{20} \Psi\e{3} - \frac{11}{10}
      \Psi\e{2,1}   - \frac{1}{3} \Psi\e{1,2}_d    +  
      \frac{2}{9} \Psi\e{1,1,1} +
      \frac{1}{4} \Psi\e{1,1,1}_c \nonumber \\
      &\qquad\quad +
      \frac{41}{150}\Psi\e{1,2}_l -
      \frac{7}{30}\Psi\e{1l,1}+
      \frac{7}{180}\Psi\e{1,1,1}_l
      \bigg]\,\beta^2t^2 \nonumber \\
     &+ {\cal O}(t^3)\label{eq:Sflow2}
\end{align}
The new terms appearing in Eq.~\eqref{eq:Sflow2} are defined as follows, where each of $\mu$, $\nu$ and $\rho$ runs independently over the four directions $\pm\hat 1$, $\pm\hat 2$ without restriction, so that backtracking terms such as $\nu=-\mu$ are included:
\begin{align}
  \Psi\e{3} =& \sum_{x,\mu,\nu,\rho}  s_x\tran s_{x+\mu  + \nu + \rho} \nonumber \\
\Psi\e{2,1} =& \sum_{x,\mu,\nu,\rho}  (s_x\tran s_{x+\mu  + \nu}) (s_x\tran s_{x + \rho}) \nonumber \\
\Psi\e{1,2}_d    =&  \sum_{x,\mu,\nu,\rho}  (s_x\tran s_{x+\mu  }) (s_{x+\nu}\tran s_{x + \rho}) \nonumber \\
 \Psi\e{1,1,1}    =&  \sum_{x,\mu,\nu,\rho}  (s_x\tran s_{x+\mu  }) (s_{x}\tran s_{x + \nu}) (s_{x}\tran s_{x + \rho}) \nonumber \\
\Psi\e{1,1,1}_c    =&  \sum_{x,\mu,\nu,\rho}  (s_x\tran s_{x+\mu  }) (s_{x+\mu}\tran s_{x + \mu + \nu}) 
(s_{x+\mu+\nu}\tran s_{x + \mu + \nu+\rho}) \nonumber \\
\Psi\e{1,2}_l  =&  \sum_{x,\mu,\nu}  (s_x\tran s_{x+\mu  }) (s_{x}\tran s_{x + \mu + \nu})\nonumber \\
\Psi\e{1l,1} =&  \sum_{x,\mu,\nu} (s_x\tran s_{x+\mu  })^2 (s_{x}\tran s_{x  + \nu})\nonumber \\
\Psi\e{1,1,1}_l =&  \sum_{x,\mu}(s_x\tran s_{x+\mu  })^3 
\end{align}

Our computer-algebra code can generate any order in the flow-time expansion. These results confirm that higher orders contain both new terms and terms already present at lower order. The new terms have increased nonlocality and increased number of spins per term. The reappearance of local terms modifies the coefficients of terms that had already appeared in lower orders. As is evident, the coefficients of the terms are functions of $\beta t$. This information may be useful for constructing nonperturbative approaches to determining these coefficients.

\subsection{Effective action}
\label{sec:EffectiveAction}

A flow action restricted to the first few orders in the flow-time expansion cannot be expected to transform the integration variables sufficiently close to a trivial distribution for a practical implementation of an improved HMC. This was demonstrated for the two-dimensional $CP(N-1)$ model, where the leading-order flow action was used~\cite{Engel:2011re}. In this study, as well as in~\cite{Luscher:2009eq}, the idea was to integrate the flow approximately with a few steps of an Euler integrator so that both the Jacobian of the discrete map and the HMC forces are computable exactly. If one attempts to use a more complex flow action and introduces a higher-accuracy integrator to better approximate the continuous flow, the numerical complexity can grow to the point that the approach becomes impractical. Computing the effective action of the transformed variables exactly would avoid these complications, but is itself impractical.
In this section, given the flow action, we derive an evolution equation for the effective action of the transformed variables and approximate its solution in a finite basis of lattice operators.

Given a flow action $S_f[s,t]$, the map $s={\cal F}_1[\sigma]$ is obtained by integrating the flow equations from $t=0$ to $t=1$, and the effective action of $\sigma$ is Eq.~\eqref{eq:exact_flowed_action} at $t=1$,
\begin{align}
    {\cal S}_{\rm eff}[\sigma]\equiv{\cal S}_0[\sigma,1]={\cal S}[{\cal F}_1[\sigma]]-\ln\det{\cal J}_1[\sigma]\,.
\end{align}
For a flow action that depends explicitly on flow time, as the perturbative solution of Sec.~\ref{sec:trivialFlow} does, the flow maps do not commute, and a closed evolution equation for the effective action must respect the order in which they are applied. Let ${\cal G}_t$ denote the map obtained by integrating the flow from flow time $t$ to flow time $1$, so that ${\cal G}_0={\cal F}_1$ and ${\cal G}_1$ is the identity, and define
\begin{align}
    {\cal B}_t[u]={\cal S}[{\cal G}_t[u]]-\ln\det D{\cal G}_t[u]\,,\qquad {\cal B}_1={\cal S}\,,\qquad {\cal B}_0={\cal S}_{\rm eff}\,.
    \label{eq:Bt}
\end{align}
The distribution $e^{-{\cal B}_t}$ is the distribution at flow time $t$ that the flow carries into the target distribution at $t=1$. The maps satisfy ${\cal G}_{t-h}={\cal G}_t\circ{\cal F}_{t\leftarrow t-h}$, where ${\cal F}_{t\leftarrow t-h}$ integrates the flow over $[t-h,t]$. Since ${\cal F}_{t\leftarrow t-h}$ rotates $u_x$ by $h\sum_a L_a\partial^a_xS_f[u,t]+{\cal O}(h^2)$, and its Jacobian determinant is $1+h\sum_{x,a}[\partial^a_x]^2S_f[u,t]+{\cal O}(h^2)$ by Eq.~\eqref{eq:jacobianODE}, one finds
\begin{align}
    \partial_t{\cal B}_t[u]=-\sum_{x,a}\partial^a_xS_f[u,t]\,\partial^a_x{\cal B}_t[u]+\sum_{x,a}[\partial^a_x]^2S_f[u,t]\,.
\end{align}
We introduce $\tau=1-t$ and $\bar{\cal S}_0[\sigma,\tau]\equiv{\cal B}_{1-\tau}[\sigma]$, so that $\bar{\cal S}_0[\sigma,0]={\cal S}[\sigma]$ is the target action and $\bar{\cal S}_0[\sigma,1]={\cal S}_{\rm eff}[\sigma]$ is the effective action of the transformed variables. Allowing for the irrelevant field-independent function $C(\tau)$ that absorbs constants, we obtain
\begin{align}
\partial_\tau\bar{\cal S}_0[\sigma,\tau]+\dot C(\tau)=
   -\sum_{x,a}[\partial_x^a]^2S_f[\sigma,1-\tau]+\sum_{x,a}\partial_x^a\bar{\cal S}_0[\sigma,\tau]\,\partial_x^aS_f[\sigma,1-\tau]\,.
   \label{eq:EffActionPDE_1}
\end{align}
The flow action enters with the reversed flow-time argument $1-\tau$: the effective action is built up starting from the last stage of the flow. For a flow action without explicit flow-time dependence, ${\bar{\cal S}}_0[\sigma,\tau]$ coincides with ${\cal S}_0[\sigma,\tau]$ of Eq.~\eqref{eq:exact_flowed_action} at every $\tau$, and Eq.~\eqref{eq:EffActionPDE_1} reduces to the equation with $S_f[\sigma]$ in place of $S_f[\sigma,1-\tau]$. In general the two agree only at $\tau=0$ and $\tau=1$.
The same equation has been derived by a different method in~\cite{Yamamura:2015kva} in the case of Yang--Mills theory. Effective lattice actions along the Wilson/gradient flow have also been extracted nonperturbatively using demon methods \cite{Kagimura:2015via}.
Equation~\eqref{eq:EffActionPDE_1} can be viewed as a gradient-flow representation of a Wilsonian RG, as discussed in \cite{Makino:2018rys,Makino:2021edc,Abe:2018zdc,Sonoda:2019ibh,Carosso:2019qpb}.
Equivalently, Eq.~\eqref{eq:EffActionPDE_1} is the continuity equation (see, for example,~\cite{Skeel2009})
\begin{align}
    \partial_\tau \rho_\tau +\nabla \cdot ( Z \rho_\tau) = 0
\end{align}
for the normalized density $\rho_\tau=e^{-\bar{\cal S}_0[\sigma,\tau]}/{\cal Z}_\tau$ of the field transported backwards from flow time $1$ to flow time $1-\tau$, with velocity field
\begin{align}
Z^a_x = -\partial^a_x S_f[\sigma,1-\tau]\,.
\label{eq:negflow}
\end{align}
The proof uses Jacobi's formula $d\ln\det J/d\tau=\Tr( J^{-1} \dot J) = \nabla \cdot Z$, as in Eq.~\eqref{eq:jacobianODE}. For the exact effective action the normalization ${\cal Z}_\tau=\int{\cal D}\sigma\,e^{-\bar{\cal S}_0[\sigma,\tau]}={\cal Z}$ does not depend on $\tau$; a $\tau$-dependent constant appears only when constants are dropped from $\bar{\cal S}_0$, and is absorbed into $C(\tau)$.

One way to solve for the effective action is to expand it in a set of linearly independent functions, thereby deriving a set of differential equations for the time-dependent coefficients of the expansion. This set of differential equations is coupled, owing to the structure of the right-hand side of Eq.~\eqref{eq:EffActionPDE_1}, and a practical application requires a truncation. We write
\begin{align}
    \bar{\cal S}_0[\sigma,\tau] = \sum_k \gamma_k(\tau) \Psi\e{[k]}[\sigma]+{\rm const}\,,
\end{align}
where the sum over $k$ runs over a restricted set of functions, and we project Eq.~\eqref{eq:EffActionPDE_1} onto this set with the connected inner product of the trivial ($t=0$) distribution, $\langle F,G\rangle=\langle FG\rangle_0-\langle F\rangle_0\langle G\rangle_0$. This gives the Galerkin equations
\begin{align}
  \sum_k\langle\Psi\e{[j]},\Psi\e{[k]}\rangle\,\dot{\gamma}_k(\tau) =&
    \Big\langle\Psi\e{[j]},-\sum_{x,a} [\partial_x^a]^2 S_f[\sigma,1-\tau]\Big\rangle \nonumber\\
    &+  \sum_k \gamma_k(\tau)\Big\langle\Psi\e{[j]},\sum_{x,a} \partial_x^a \Psi\e{[k]}  \partial_x^a S_f[\sigma,1-\tau] \Big\rangle \,,
   \label{eq:EffActionPDE_2}
\end{align}
with the initial condition $\bar{\cal S}_0[\sigma,0]={\cal S}[\sigma]$. If the $\Psi\e{[k]}$ are eigenfunctions of the Lie Laplacian with distinct eigenvalues, the matrix on the left-hand side is diagonal.

To illustrate the procedure, we use the four eigenfunctions of Sec.~\ref{sec:trivialFlow},
\begin{align}
     \bar{\cal S}_0[\sigma,\tau] = \gamma^{(0)}(\tau) \Psi\e{0}[\sigma] + \gamma\e{2}(\tau) \Psi\e{2}[\sigma] + \gamma\e{1,1}_l(\tau) \Psi\e{1,1}_l[\sigma]  + \gamma\e{1,1}(\tau)\Psi\e{1,1}[\sigma]
     \,,
\end{align}
with $\gamma^{(0)}(0)=-\beta/2$ and $\gamma\e{2}(0)=\gamma\e{1,1}_l(0)=\gamma\e{1,1}(0)=0$. They are orthogonal, with $\langle\Psi\e{0},\Psi\e{0}\rangle=\frac83|\Lambda|$, $\langle\Psi\e{2},\Psi\e{2}\rangle=\frac{40}{3}|\Lambda|$, $\langle\Psi\e{1,1}_l,\Psi\e{1,1}_l\rangle=\frac{32}{45}|\Lambda|$ and $\langle\Psi\e{1,1},\Psi\e{1,1}\rangle=\frac{16}{9}|\Lambda|$.
For the zeroth-order flow action
\begin{align}
     S_f[s] = \frac{\beta}{4 C_F} \Psi\e{0}[s]\,,
\end{align}
the Laplacian term is $-\sum_{x,a}[\partial^a_x]^2S_f=\frac{\beta}{2}\Psi\e{0}$, and the contraction of $\Psi\e{0}$ with the flow action was computed in Eq.~\eqref{eq:S0contrFlow}:
\begin{align}
    \gamma^{(0)}(\tau) \sum_{x,a}\partial_x^a \Psi\e{0} \partial_x^a S_f=&  \frac{\gamma^{(0)}(\tau) \beta}{C_F} \left[ \frac{C_F}{1+ C_F} \Psi\e{2}
    - \Psi\e{1,1}  -   \Psi\e{1,1}_l
    \right] + {\rm const}\,.
\end{align}
The contractions of the other three functions with the flow action do not lie in the span of the four functions. One finds, for example,
\begin{align}
    \sum_{x,a}\partial_x^a \Psi\e{2} \partial_x^a \Psi\e{0}=4\left(\Psi\e{3}-\Psi\e{2,1}\right),\qquad
    \sum_{x,a}\partial_x^a \Psi\e{1,1}_l \partial_x^a \Psi\e{0}=8\left(\Psi\e{1,2}_l-\Psi\e{1l,1}\right),
    \label{eq:closure}
\end{align}
with the functions defined in Sec.~\ref{sec:trivialFlow}. Their projections onto the four functions are nonzero only along $\Psi\e{0}$:
\begin{align}
    \gamma\e{2} \sum_{x,a}\partial_x^a \Psi\e{2} \partial_x^a S_f\;\to&\;\frac{5\beta}{3}\gamma\e{2}\,\Psi\e{0}\,,\nonumber\\
    \gamma\e{1,1}_l \sum_{x,a}\partial_x^a \Psi\e{1,1}_l \partial_x^a S_f\;\to&\;\frac{2\beta}{5}\gamma\e{1,1}_l\,\Psi\e{0}\,,\nonumber\\
    \gamma\e{1,1} \sum_{x,a}\partial_x^a \Psi\e{1,1} \partial_x^a S_f\;\to&\;\frac{5\beta}{3}\gamma\e{1,1}\,\Psi\e{0}\,.
\end{align}
Setting $C_F=2$, we obtain
\begin{align}
    \dot\gamma^{(0)} =& \frac{\beta}{2} + \frac{5\beta}{3}\gamma\e{2}+\frac{2\beta}{5}\gamma\e{1,1}_l+\frac{5\beta}{3}\gamma\e{1,1}\,,  \nonumber\\
    \dot\gamma\e{2} =& \frac{\beta}{3}\gamma^{(0)}\,,\qquad
    \dot\gamma\e{1,1}_l = - \frac{\beta}{2}\gamma^{(0)}\,,  \qquad
    \dot\gamma\e{1,1} = - \frac{\beta}{2}\gamma^{(0)}\,.
    \label{eq:gamma0}
\end{align}
It follows that $\ddot\gamma^{(0)}=-\frac{43}{90}\beta^2\gamma^{(0)}$, and with $\Omega=\beta\sqrt{43/90}$ the solution is
\begin{align}
    \gamma^{(0)}(\tau) &= -\frac{\beta}{2}\cos(\Omega\tau) + \frac{3\sqrt{430}}{86}\sin(\Omega\tau)\,,\nonumber\\
    \gamma^{(2)}(\tau) &= \frac{15}{43}\left(1-\cos(\Omega\tau)\right) - \frac{\sqrt{430}}{86}\,\beta\sin(\Omega\tau)\,,\qquad
    \gamma^{(1,1)}_l(\tau)=\gamma^{(1,1)}(\tau) = -\frac{3}{2}\gamma^{(2)}(\tau)\,.
\end{align}

\begin{figure}[t]
    \centering
    \includegraphics[width=\textwidth]{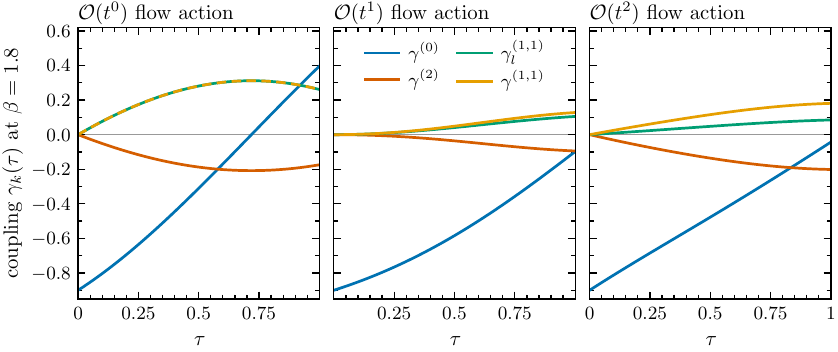}
    \caption{Couplings $\gamma^{(0)}$, $\gamma^{(2)}$, $\gamma^{(1,1)}_l$ and $\gamma^{(1,1)}$ of the four-function truncation of the effective action, Eqs.~\eqref{eq:gamma0}--\eqref{eq:gamma2}, at $\beta=1.8$, for the flow action truncated at ${\cal O}(t^0)$ (left), ${\cal O}(t^1)$ (center) and ${\cal O}(t^2)$ (right). The target action corresponds to $\tau=0$ and the effective action of the transformed variables to $\tau=1$. In the left panel $\gamma^{(1,1)}_l=\gamma^{(1,1)}$.}
    \label{fig:coupling_flow}
\end{figure}

For the first-order flow action $S_f=S_f\e{0}+t\,S_f\e{1}$ of Sec.~\ref{sec:trivialFlow}, the flow action enters Eq.~\eqref{eq:EffActionPDE_2} with $t=1-\tau$. Both the Laplacian of $S_f\e{1}$ and its contractions with the four functions contribute, and the right-hand sides of Eq.~\eqref{eq:gamma0} acquire the additional terms
\begin{align}
    \delta\dot\gamma^{(0)} =& \frac{7}{45}\beta^2(1-\tau)\,\gamma^{(0)}\,,\nonumber\\
    \delta\dot\gamma\e{2} =& \beta^2(1-\tau)\left[\frac{1}{6}+\frac{8}{15}\gamma\e{2}-\frac{2}{25}\gamma\e{1,1}_l-\frac{5}{27}\gamma\e{1,1}\right]\,,\nonumber\\
    \delta\dot\gamma\e{1,1}_l =& \beta^2(1-\tau)\left[-\frac{1}{4}-\frac{1}{6}\gamma\e{2}-\frac{1}{21}\gamma\e{1,1}_l+\frac{13}{90}\gamma\e{1,1}\right]\,,\nonumber\\
    \delta\dot\gamma\e{1,1} =& \beta^2(1-\tau)\left[-\frac{1}{4}-\frac{7}{18}\gamma\e{2}+\frac{19}{75}\gamma\e{1,1}_l+\frac{26}{135}\gamma\e{1,1}\right]\,.
    \label{eq:gamma1}
\end{align}
The second-order flow action $S_f=S_f\e{0}+t\,S_f\e{1}+t^2S_f\e{2}$ adds
\begin{align}
    \delta\dot\gamma^{(0)} =& \beta^3(1-\tau)^2\left[\frac{7}{90}+\frac{83}{270}\gamma\e{2}+\frac{289}{13125}\gamma\e{1,1}_l+\frac{32947}{202500}\gamma\e{1,1}\right]\,,\nonumber\\
    \delta\dot\gamma\e{2} =& \frac{83}{1350}\beta^3(1-\tau)^2\gamma^{(0)}\,,\nonumber\\
    \delta\dot\gamma\e{1,1}_l =& -\frac{551}{31500}\beta^3(1-\tau)^2\gamma^{(0)}\,,\qquad
    \delta\dot\gamma\e{1,1} = -\frac{6647}{135000}\beta^3(1-\tau)^2\gamma^{(0)}\,.
    \label{eq:gamma2}
\end{align}
These systems are integrated numerically. The resulting couplings are shown in Fig.~\ref{fig:coupling_flow} for $\beta = 1.8$.

Equation~\eqref{eq:closure} shows that the four-function truncation is not closed: the $\Psi\e{0}$ couplings generated by $\gamma\e{2}$, $\gamma\e{1,1}_l$ and $\gamma\e{1,1}$ arise from projecting functions that lie outside the truncated space. Enlarging the space to the twelve functions $\Psi\e{0}$, $\Psi\e{2}$, $\Psi\e{1,1}_l$, $\Psi\e{1,1}$, $\Psi\e{3}$, $\Psi\e{2,1}$, $\Psi\e{1,2}_d$, $\Psi\e{1,1,1}$, $\Psi\e{1,1,1}_c$, $\Psi\e{1,2}_l$, $\Psi\e{1l,1}$, $\Psi\e{1,1,1}_l$ of Sec.~\ref{sec:trivialFlow}, which are not mutually orthogonal, closes the contractions of Eq.~\eqref{eq:closure}. In this space, the zeroth-order flow action gives $\gamma^{(0)}(\tau)=-\frac{\beta}{2}(1-\tau)$ exactly. For the flow actions truncated at ${\cal O}(t^0)$, ${\cal O}(t^1)$ and ${\cal O}(t^2)$, the twelve-function system gives an effective action whose expansion in $\beta$ starts at ${\cal O}(\beta^2)$, ${\cal O}(\beta^3)$ and ${\cal O}(\beta^4)$, respectively, so the target action is removed order by order, as required by the construction of Sec.~\ref{sec:trivialFlow}. Through ${\cal O}(\beta^3)$ all contractions lie in the span of the twelve functions, and the truncated solution coincides with the expansion of the exact effective action. At $\beta=1.8$, however, the couplings of the four functions change substantially when the space is enlarged to twelve functions, so the curves of Fig.~\ref{fig:coupling_flow} illustrate the procedure and are not converged in the truncation. Section~\ref{sec:Numerics} does not use these truncated coefficients; there the approximate effective action is determined nonperturbatively, as described in Sec.~\ref{sec:adjoint}. Section~\ref{sec:Numerics} also assesses the truncation of the flow action through the standard deviation $\sigma_{\Delta S}$ of the exact action mismatch.
One could also adopt other flow actions with a less severe truncation and a time dependence chosen to yield an effective action with reduced critical slowing down that is well approximated with a reasonable number of terms. The gradient flow of the two-dimensional $\mathrm{O}(N)$ nonlinear sigma model itself has been studied in perturbation theory and at large $N$~\cite{Makino:2014sta,Makino:2014cxa,Aoki:2014dxa}.

If one determines a flow action whose effective action can be approximated to reasonable precision with a manageable number of terms, an improved simulation algorithm for the target action may be possible. Such an algorithm proceeds with the following steps, assuming that the map from the new variables $\sigma$ to the original variables $s$ is $\mathcal{F}[\sigma]=s$:
\begin{enumerate}
    \item Given a configuration $s$, construct $\sigma=\mathcal{F}^{-1}[s]$ and compute the exact effective action $\mathcal{S}_0$ given by Eq.~\eqref{eq:exact_flowed_action}.
    \item Using an approximate effective action $\mathcal{S}_0^{\rm ap}$, for example the truncated solution constructed above or the nonperturbative determination of Sec.~\ref{sec:adjoint}, perform Hamiltonian evolution with normally distributed momenta $p$ to obtain $[p',\sigma']$.
    \item Compute $s'=\mathcal{F}[\sigma']$ and the exact  effective action $\mathcal{S}_0$ using Eq.~\eqref{eq:exact_flowed_action}.
    \item Perform a Metropolis Accept/Reject step using the exact effective actions computed before and after the Hamiltonian evolution by integrating the flow equations that define the map $\mathcal{F}$.
\end{enumerate}

This is a variant of the algorithm proposed by L\"uscher for the gauge theory~\cite{Luscher:2009eq}: there the molecular-dynamics force is the exact derivative of the effective action of the discretized map, whereas here it is evaluated approximately from $\mathcal{S}_0^{\rm ap}$ and exactness is restored by the accept/reject step.
The algorithm is exact; its effectiveness depends on the acceptance rate in the last step and the ``triviality'' of the effective action. High acceptance requires the approximate effective action to be sufficiently close to the true effective action resulting from the flow that defines the map $\mathcal{F}$. The computational cost of such an algorithm is the cost of solving the flow equation twice plus the cost of the Hamiltonian evolution. The cost of the Hamiltonian evolution increases as the approximate effective action becomes more complex. However, it will not be larger than the cost of computing the exact force derived from the exact effective action, where each force evaluation requires solving the flow equation, as shown below. The ``triviality'' of the effective action is also important: a small approximate force allows faster mixing in phase space.

The algorithm is exact for any choice of $\mathcal{S}_0^{\rm ap}$: the molecular-dynamics map is reversible and volume preserving for any force, and the accept/reject step uses the exact effective action. The approximation enters only through the mismatch $\Delta S=\mathcal{S}_0[\sigma,1]-\mathcal{S}_0^{\rm ap}[\sigma]$. Since $\langle e^{-\delta H}\rangle=1$ for such a proposal, an approximately Gaussian energy change $\delta H$ has mean ${\rm Var}(\delta H)/2$. If the molecular dynamics conserves its own Hamiltonian accurately and the trajectory decorrelates $\Delta S$, then ${\rm Var}(\delta H)\approx2\sigma_{\Delta S}^2$, where $\sigma_{\Delta S}$ is the standard deviation of $\Delta S$, and the acceptance is ${\rm erfc}(\sigma_{\Delta S}/2)$. The same quantity controls reweighting: configurations drawn from $e^{-\mathcal{S}_0^{\rm ap}}$ and reweighted by $e^{-\Delta S}$ give $N_{\rm eff}/N\approx e^{-\sigma_{\Delta S}^2}$. The coefficients of $\mathcal{S}_0^{\rm ap}$ need not be taken from the truncated evolution equation. Section~\ref{sec:adjoint} determines them nonperturbatively by minimizing a Kullback--Leibler divergence, and Sec.~\ref{sec:Numerics} uses that determination.

\subsection{Nonperturbative flow action}
\label{sec:adjoint}

The perturbative construction of the flow action is unlikely to be adequate for an efficient algorithm and therefore nonperturbative methods are required. A variational approximation can be performed by introducing a flow action that depends on certain parameters. The precise parametrization is not important at this stage. In this approach, the optimal flow action can be found by minimizing an appropriately defined functional. 
The Kullback--Leibler divergence defines a measure of distance between distributions and can therefore be used to optimize the flow action. 
The Kullback--Leibler divergence is defined as
\begin{align}
    D_{KL}[q|p] = \int \mathcal{D} s\, q[s] \ln\left(\frac{q[s]}{p[s]}\right)
\end{align}
where $p[s]$ and $q[s]$ are the two distributions being compared. It is non-negative and vanishes if and only if the two distributions agree almost everywhere. 
In our case we take
\begin{align}
    p[s] {\cal D}s =& \frac{e^{-{\cal S}[s]}}{\cal Z} {\cal D}s \nonumber\\
    q[s] {\cal D}s =& \frac{e^{-\ln \det {\cal J}_1[s]}}{{\cal Z}_q} {\cal D}s = \frac{1}{{\cal Z}_q} {\cal D} \sigma
\end{align}
Here the flow defines the forward map from the trivial variables to the target
variables, $\sigma\mapsto s=s_1[\sigma]$, and ${\cal J}_1[\sigma]$ denotes the
Jacobian matrix of this map.  Equivalently, ${\cal D}s=\det{\cal J}_1[\sigma]\,
{\cal D}\sigma$, so the induced density $q[s]$ with respect to the ${\cal D}s$
measure contains the inverse determinant which can be expressed in terms of the $s$ variables. The distribution for $\sigma$ is taken
to be trivial. With these definitions we obtain
\begin{align}
    D_{KL}[q|p] = \frac{1}{{\cal Z}_q}\int \mathcal{D} \sigma\, \left\{{\cal S}[s_1[\sigma]] - \ln\det{\cal J}_1[\sigma]\right\}  + \ln \frac{\cal Z}{{\cal Z}_q}
    \label{eq:KL-div}
\end{align}
The constants are independent of the transformation parameters and therefore are not needed for the computation of the derivatives with respect to the flow-action parameters used in the optimization process. 
These derivatives are
\begin{align}
    \frac{\partial D_{KL}[q|p] }{\partial c_i} = \frac{1}{{\cal Z}_q}\int \mathcal{D} \sigma\, \frac{\partial \ell[s_1[\sigma]]}{\partial c_i}
\end{align}
where $c_i$ is the $i$-th parameter of the flow action, and
\begin{align}
    \ell[s_1]= {\cal S}[s_1]- \ln\det {\cal J}_1 = {\cal S}[s_1] - \int_0^1dt'\;\sum_{x,a} [\partial^a_x]^2 S_f[s_{t'},t']\, .
\end{align}
We compute them stochastically by sampling the trivial distribution of $\sigma$ and evaluating $\partial \ell[s_1[\sigma]]/{\partial c_i}$.

Another type of derivative we may need to evaluate is the derivative of the effective action with respect to the $\sigma$ variables. 
If the optimized flow does not trivialize the theory, and if the resulting effective-action parametrization is not sufficiently accurate to implement the algorithm outlined in the previous section, one must perform an HMC simulation using the effective action. This is the algorithm originally proposed by L\"uscher for the gauge theory~\cite{Luscher:2009eq}.
As noted above, the effective action for the $\sigma$ variables is $\ell[s_1[\sigma]]$; therefore the force needed in the HMC simulation is
\begin{align}
    f^a_x = - \hat\partial^a_x \ell[s_1[\sigma]]\,,
\end{align}
where $\hat\partial^a_x$ are the Lie derivatives with respect to $\sigma$.

Both desired derivatives of $\ell[s]$ can be computed using adjoint sensitivity~\cite{Cao2003AdjointSA}. For gradient flows on Lie groups, the adjoint-state method has been used to learn the flow-action coefficients of $\mathrm{SU}(N)$ gauge theories~\cite{Bacchio:2022vje}, where the adjoint state also yields the HMC force in the trivialized variables, and to train continuous flows~\cite{Gerdes:2024rjk}. In Ref.~\cite{Luscher:2009eq} the HMC force for the Euler-integrated Wilson flow is obtained by propagating derivatives backward through the integration steps, a discrete analogue of the same construction.

\subsubsection{Adjoint sensitivity}

The derivatives of $\ell[s]$ can be constructed using the Lagrange-multiplier method.
We define
\begin{align}
    \tilde\ell = \ell + \int_0^1 dt \sum_x \lambda\tran_x \left[ \dot s_x - \sum_a L_a \partial_x^aS_f[s] s_x \right]\,,
    \label{eq:augLoss}
\end{align}
where $s_x$ is understood to be time dependent and the Lagrange multiplier $\lambda_x$ is a three-component vector in $\mathbb{R}^3$ per site $x$. A functional of this type, with a Lagrange-multiplier field enforcing the flow equation, was used by L\"uscher and Weisz~\cite{Luscher:2011bx} (following Ref.~\cite{Zinn-Justin:1987its}) to derive the Feynman rules for correlation functions of gradient-flowed gauge fields. 
On trajectories satisfying the flow equation, Eq.~\eqref{eq:flow_equation_general}, the constraint term vanishes and the above functional is identical to the original $\ell$. Therefore, their derivatives agree on the flow trajectory.  

The Lie derivative $\partial^a_x$ with respect to $s(t)$ at a flow time $t$ is related to $\hat\partial^a_x$ via the Jacobian matrix
\begin{align}
    \hat\partial^a_x = \sum_{b,y} {\cal J}^{b,a}_t(y,x)\partial^b_y
\end{align}
while a similar expression holds for the total derivative with respect to a parameter $c_i$
\begin{align}
    d_i \equiv \frac{d}{dc_i} = \sum_{a,y} \omega^a_i(y,t) \partial^a_y  + \frac{\partial}{\partial c_i}\,.
\end{align}
This can be understood by introducing variations of the spin field that preserve the norm and defining the left derivative with respect to $c_i$ as
\begin{align}
    \frac{\partial s_x(t)}{\partial c_i} = \lim_{\delta c_i\rightarrow 0 }\frac{e^{\sum_a L_a \omega^a_i(x,t) \delta c_i} -1 }{\delta c_i} s_x = \sum_a L_a \omega^a_i(x,t) s_x(t)\,.
\end{align}
Both the Jacobian matrix and $\omega^a_i(y,t)$ are unknown and depend on the flow action, but we do not need to determine them in order to compute the desired derivatives. The Lagrange multiplier can be chosen so that all terms containing these matrices are eliminated from the derivatives. To unify the notation, we introduce the matrix
\begin{align}
    M^b_k(y,t) \equiv  {\cal J}^{b,a}_t(y,x)
\end{align}
where $k$ is a combined index for $x$ and $a$, or
\begin{align}
    M^b_k(y,t) \equiv \omega^b_i(y,t)
\end{align}
where $k$ denotes the index $i$.
With this notation, the total derivative $\delta_k$ is
\begin{align}
    \delta_k \equiv  \sum_{a,y} M^a_k(y,t) \partial^a_y  + \partial_k\,.
\end{align}
In this notation, $\delta_k$ is either a Lie derivative $\hat\partial^a_x $ or a parameter derivative $d_i$ and  $\partial_k$ is understood to vanish when $\delta_k$ denotes a Lie derivative.  This unified notation leads to
 \begin{align}
     \delta_k \tilde\ell  =& \sum_{x,b} M^b_k(x,1) \left[ \partial_x^b {\cal S}   
     + \xi^b_x(1) \right] -  \sum_{b,x} \xi^b_x(0) M^b_k(x,0) + \partial_k {\cal S} - \nonumber \\
     &-  \int_0^1 dt \,\, \sum_{x,b}M^b_k(x,t)\left\{\dot\xi^b_x(t) + \partial_x^b\left[ \sum_{y,a} [\partial_y^a]^2 S_f[s]\right] \right. \nonumber \\
     &\qquad \left. {}-  \sum_{a,c}\epsilon_{abc} \xi^a_x(t) \partial^c_x S_f[s] + \sum_{c,y} \xi^c_y(t) \partial_x^b \partial_y^c S_f[s]  \right\} -\nonumber \\
    & - \int_0^1 dt \,\,  \sum_{y,b}  \partial_k  \left[(\partial_y^b)^2S_f[s]\right]  - \int_0^1 dt \sum_{b,x} \xi^b_x(t) \partial_k \partial_x^b S_f[s] 
 \end{align}
 where
\begin{align}
    \xi^a_x(t) = \lambda\tran_x L_a s_x(t)\,.
\end{align}
All terms proportional to the unknown $ M^b_k(x,t)$ matrices can be eliminated by setting
\begin{align}
    \dot\xi^a_x +\sum_{b,y} \partial_x^a \partial_y^b S_f[s] \xi^b_y   - \sum_{b,c}   \epsilon_{abc} \partial^b_x S_f[s] \xi^c_x  +
     \partial_x^a\left[ \sum_{y,b} [\partial_y^b]^2 S_f[s]\right] =0
     \label{eq:adjointflow}
 \end{align} 
and
 \begin{align}
 \partial^a_x{\cal S} + \xi^a_x(1) = 0
 \label{eq:adjointBC}
 \end{align}
 which provides the boundary condition for solving Eq.~\eqref{eq:adjointflow}. 
 Using 
 \begin{align}
     [\partial^a_x,\partial^b_y] = - \epsilon_{abc} \partial^c_x \delta_{x,y}
 \end{align}
 we get
\begin{align}
    \dot\xi^a_x +\sum_{b,y} \xi^b_y \Big(\partial_y^b \partial_x^a S_f[s]\Big)     +
     \partial_x^a\left[ \sum_{y,b} [\partial_y^b]^2 S_f[s]\right] =0
     \label{eq:adjointflow2}
 \end{align} 
 This equation, together with the boundary condition Eq.~\eqref{eq:adjointBC}, can be solved with standard Runge--Kutta methods since it is an ordinary differential equation for $\xi^a_x$. Since $\partial_k {\cal S} = 0$, because ${\cal S}$ does not depend explicitly on the flow initial condition or flow parameters, the derivative of $\tilde\ell$ is
 \begin{align}
     \delta_k \tilde\ell = - \int_0^1 dt \,\,  \sum_{y,b}  \partial_k  \left[(\partial_y^b)^2S_f[s]\right] -  \sum_{b,y} \xi^b_y(0) M^b_k(y,0)
     -  \int_0^1 dt \sum_{b,y} \xi^b_y(t) \partial_k \partial_y^b S_f[s] 
 \end{align}

 We now calculate both types of desired derivatives. First, the HMC force for the effective action in the $\sigma$ variables is obtained by noting that all terms containing $\partial_k S_f$ vanish because $S_f$ has no explicit dependence on the $\sigma$ field. Therefore, 
 \begin{align}
     f^a(x) = -\hat\partial^a_x \ell = -\delta_{a,x} \tilde\ell = \sum_{b,y} \xi^b(y,0) M^b_{a,x}(y,0)
 \end{align}
 
However, $ M^b_{a,x}(y,0) = \mathcal{J}_0^{b,a}(y,x) = \delta_{a,b}\delta_{x,y}$ (here we use $k\equiv{(a,x)}$), resulting in

 \begin{align}
     f^a(x) = -\hat\partial^a_x \ell = -\delta_{a,x} \tilde\ell = \xi^a(x,0)
 \end{align}
 which can be found by solving the ODE in Eq.~\eqref{eq:adjointflow} for $\xi^a$ with the prescribed boundary condition at $t=1$ [Eq.~\eqref{eq:adjointBC}].
 
 Similarly, one can obtain derivatives with respect to the flow parameters at fixed initial condition $\sigma$. 
 In this case $M^b_k(x,0)=0$ because $\sigma$ does not depend on the flow parameters $c_k$ and
 \begin{align}
     \pder{c_k}\ell =  - \int_0^1 dt \,\,  \sum_{y,b}  \partial_k  \left[(\partial_y^b)^2S_f[s]\right]  
     -  \int_0^1 dt \sum_{b,y} \xi^b(y) \partial_k \partial_y^b S_f[s] \,.
     \label{eq:paramGrad}
 \end{align}
 which can be computed by direct integration once $\xi^a_x$ is known. 
 
\subsubsection{Flow and effective action parametrization}
With the derivative methodology outlined here, we set up a fully nonperturbative tuning of a truncated flow action. Motivated by the structure of the perturbative flow action, we search for the flow action that minimizes the KL divergence in Eq.~\eqref{eq:KL-div}, parametrized as
\begin{align}
    S_f[s,t] = \beta\sum_k g\e{k}(t)\, \Phi_k[s]
\end{align}
where the $\Phi_k[s]$, labeled by an integer $k$, are functions obtained in the perturbative evaluation of the flow action, such as those in Eq.~\eqref{eq:Sflow2}. The functions $g\e{k}(t)$ are unknown; in the perturbative solution they depend on $\beta$ and $t$ only through $\beta t$. They can themselves be expressed as a series of orthogonal polynomials. For example, we may choose Chebyshev polynomials $T_n$ so that
\begin{align}
g\e{k}(t) = \sum_{n=0}^{N_{\rm max}} c\e{k}_n\, T_n(2 t - 1)
\end{align}
where $ c\e{k}_n $ are functions of $\beta$ determined by the minimization of the KL divergence using a stochastic evaluation of the gradient as outlined in the previous section.
A better parametrization may be achieved using Laguerre or Laguerre--Sobolev polynomials, which naturally impose that the $g\e{k}$ are functions of $\beta t$ and that their domain is the nonnegative real axis.

Once the flow action is fixed, the approximate effective action ${\cal S}_0^{\rm ap}$ required by the algorithm of Sec.~\ref{sec:EffectiveAction} can also be determined nonperturbatively. We expand it in translation-invariant functions $\Psi\e{[k]}$, for example the functions used to parametrize the flow action, possibly supplemented by further functions:
\begin{align}
     {\cal S}_{0}[\sigma,1] =& {\cal S}[{\cal F}_1[\sigma]]-  \int_0^1 dt' \,\, \sum_{x,a} [\partial_x^a]^2 S_f[{\cal F}_{t'}[\sigma],t'] ={\cal S}[{\cal F}_1[\sigma]]-  
     \ln\det{\cal J}_1[\sigma]\nonumber \\
     \approx& \sum_k \gamma\e{k}\Psi\e{[k]}[\sigma]\equiv{\cal S}_0^{\rm ap}[\sigma]\,.
\end{align}
The coefficients $\gamma\e{k}$ can be determined 
by minimizing the KL divergence 
\begin{align}
    D_{KL}(q_0| p) = \int {\mathcal{D}}\,\sigma\, q_0[\sigma]\ln\left(\frac{q_0[\sigma]}{p[\sigma]}\right)
    \label{eq:KLsur}
\end{align}
with 
\begin{align}
    q_0[\sigma] =& \frac{e^{- {\cal S}_0^{\rm ap}[\sigma]}}{{\cal Z}_0}\,,\nonumber\\
    p[\sigma] =&  \frac{e^{- {\cal S}[{\cal F}_1[\sigma]]+  \ln\det{\cal J}_1[\sigma] }} {{\cal Z}}\,,
\end{align}
resulting in
\begin{align}
    D_{KL}(q_0| p) = \int {\mathcal{D}}\,\sigma\,  \frac{e^{- {\cal S}_0^{\rm ap}[\sigma]}}{{\cal Z}_0}\left( 
     {\cal S}[{\cal F}_1[\sigma]]-  \ln\det{\cal J}_1[\sigma]  - {\cal S}_0^{\rm ap}[\sigma]\right) - \ln\frac{{\cal Z}_0}{\cal Z}\,.
\end{align}
The computation of the derivatives with respect to $\gamma\e{k}$ is
straightforward, and one obtains
\begin{align}
    \pder{\gamma\e{k}} D_{KL}(q_0| p) =  -\langle (\Psi\e{[k]} - \overline{\Psi\e{[k]} })(\Delta - \overline{\Delta })\rangle =  -{\rm Cov}(\Psi\e{[k]} ,\Delta)\,,
    \label{eq:KLgrad}
\end{align}
where $\Delta \equiv  {\cal S}[{\cal F}_1[\sigma]]-  \ln\det{\cal J}_1[\sigma] - {\cal S}_0^{\rm ap}[\sigma]$. Above, $\overline A = \langle A\rangle$, and $\langle\cdot\rangle$ denotes the average over the $q_0$ distribution. Note that when the parametrization exactly matches the effective action, $\Delta=0$, and hence the derivatives vanish. At any stationary point ${\rm Cov}(\Psi\e{[k]},\Delta)=0$ for every $k$: the mismatch is orthogonal to all functions of the expansion with respect to the $q_0$ average. This is the counterpart of the Galerkin projection of Eq.~\eqref{eq:EffActionPDE_2}, applied to the exact effective action at $\tau=1$ rather than to its evolution equation, so that no truncation error is carried along the flow. The derivatives are estimated stochastically via Monte Carlo sampling of the $q_0$ distribution, which should be close to trivial if the trivializing map is well approximated. Therefore, no critical slowing down should occur.
Certain optimizers may require the Hessian matrix as well. It can be easily computed as
\begin{align}
    H_{lk} =\pder{\gamma\e{l}}\pder{\gamma\e{k}}  D_{KL}(q_0| p) =&
    \langle (\Psi\e{[k]} - \overline{\Psi\e{[k]} })(\Psi\e{[l]} - \overline{\Psi\e{[l]}}  )\rangle \nonumber\\
    &+\langle (\Psi\e{[k]} - \overline{\Psi\e{[k]} })(\Delta - \overline{\Delta })(\Psi\e{[l]}    -\overline{\Psi\e{[l]} })\rangle \,.
\end{align}
At a minimum where $\Delta = 0$, the third-cumulant term vanishes and the Hessian reduces to the covariance matrix of the $\Psi\e{[k]}$, which is positive semidefinite (definite if the $\Psi\e{[k]}$ are linearly independent modulo constants).

\section{Numerical Experiments}
\label{sec:Numerics}

\newcommand{\Err}{\sigma_{\Delta S}}

We now test the constructions of Sec.~\ref{sec:O3model} numerically. Section~\ref{sec:num_pert} measures the quality of the truncated flows as stand-alone maps; Sec.~\ref{sec:num_thmc} uses the second-order map in transformed HMC and compares autocorrelation times and cost with direct HMC; Sec.~\ref{sec:num_np} tunes the flow-action coefficients nonperturbatively; Sec.~\ref{sec:num_sur} constructs approximate effective actions for the algorithm of Sec.~\ref{sec:EffectiveAction}; and Sec.~\ref{sec:num_mts} tests them as force surrogates for multi-timestep HMC.

\subsection{Perturbative flows}
\label{sec:num_pert}

We first test the truncated flows constructed above as stand-alone maps, extending the tests of Ref.~\cite{Chamness:2024fob}. Fields of uniformly distributed unit vectors $\sigma$ are transported to $t=1$, and Eq.~\eqref{eq:exact_flowed_action} at $t=1$ gives ${\cal S}_{0}[\sigma,1]$ for each sample. For an exact trivializing map this quantity does not depend on $\sigma$; the quality of an approximate map is therefore measured by the standard deviation $\sigma_{\Delta S}$ of $\Delta S= {\cal S}_{0}[\sigma,1]$ over trivially distributed $\sigma$, which is the mismatch defined in Sec.~\ref{sec:EffectiveAction} for the trivial choice ${\cal S}_0^{\rm ap}=0$. We label the truncations of the small flow-time expansion of Sec.~\ref{sec:trivialFlow} by their order $n$: the flow action of order $n$ is $\sum_{k=0}^{n}t^kS_f^{(k)}$, and we test $n=0$, 1 and 2.

\begin{figure}[t]
\centering
  \includegraphics[width=\linewidth]{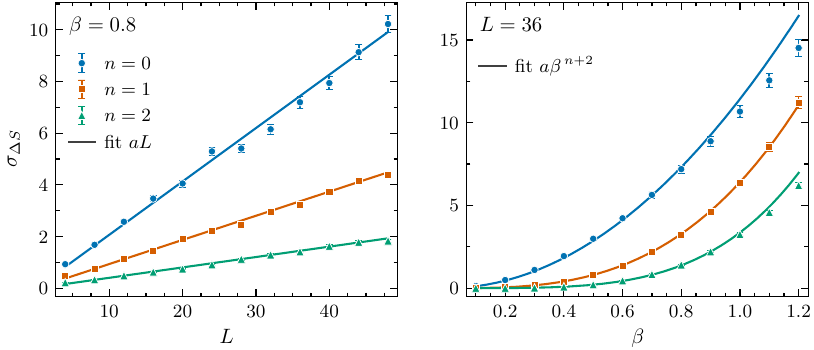}
  \caption{ Accuracy of the approximate trivializing maps, quantified by the standard 
  deviation $\sigma_{\Delta S}$ of the difference $\Delta S$ between the true and 
  approximate actions. 
  \textbf{Left:} $\sigma_{\Delta S}$ as a function of lattice size $L$ at fixed 
  $\beta=0.8$. 
  \textbf{Right:} $\sigma_{\Delta S}$ as a function of $\beta$ at fixed lattice size $L=36$. 
  Results are shown for the flow action truncated at order $n=0$, 1 and 2 in the flow time, with jackknife errors.
  Lines: weighted one-parameter fits $\sigma_{\Delta S}=aL$ (left) and $\sigma_{\Delta S}=a\beta^{n+2}$ (right).
  Higher-order truncations consistently reduce the error, but the scaling with 
  the lattice size remains approximately linear, indicating exponential degradation 
  with volume. The growth with $\beta$ is polynomial at small $\beta$, $\sigma_{\Delta S}\propto\beta^{n+2}$.
  \label{fig:test1}}
\end{figure}

Each estimate of $\Err$ uses 512 trivially distributed configurations, transported with a fourth-order Runge--Kutta--Munthe-Kaas integrator~\cite{MUNTHEKAAS1999115} at fixed step size $\delta t=0.02$; statistical errors are obtained with the jackknife. We study how $\Err$ depends on the lattice volume $|\Lambda|$ and on the coupling $\beta$ of the target action.
At fixed $\beta = 0.8$ (Fig.~\ref{fig:test1}, left), $\Err$ grows approximately linearly with the lattice side $L=|\Lambda|^{1/2}$ at each order of the flow-time expansion, and each additional order lowers $\Err$ at every $L$. This growth is expected for a local map: $\Delta S$ is a sum of local contributions, so for $L\gg\xi$ its variance is extensive and $\Err\propto\sqrt{|\Lambda|}=L$. The overlap of a fixed local map with the target distribution therefore degrades exponentially with the volume~\cite{Finkenrath:2022ogg,Abbott:2022zsh}; for approximately Gaussian $\Delta S$, $N_{\rm eff}/N\approx e^{-\Err^2}\propto e^{-cL^2}$.
The right panel of Fig.~\ref{fig:test1} shows the dependence on $\beta$ at fixed $L=36$. Although the expansion is in the flow time on $0 \le t \le 1$, each further order brings one more contraction with $\mathcal{S}$ and hence one more power of $\beta$. A truncation at order $n$ therefore leaves a mismatch $\Err\sim\beta^{n+2}$ at small $\beta$, i.e.\ $\beta^2$, $\beta^3$ and $\beta^4$ at orders 0, 1 and 2.

A good measure of the effectiveness of the trivializing flow as a simulation algorithm is the \emph{effective sample size} after reweighting, denoted ESS (reweight). This standard measure estimates how many independent, \emph{unweighted} draws the weighted sample is worth.
For importance weights \(w_i \propto p(x_i)/q(x_i)\) with normalized \(\tilde w_i = w_i/\sum_{j=1}^N w_j\), the classical ``Kish'' ESS used in importance sampling  is
\begin{equation}
N_{\mathrm{eff}}
=\frac{\Big(\sum_{i=1}^N w_i\Big)^2}{\sum_{i=1}^N w_i^2}
=\frac{1}{\sum_{i=1}^N \tilde w_i^{\,2}}\,.
\label{eq:ess}
\end{equation}

\begin{figure}[t]
\centering
  \includegraphics[width=\linewidth]{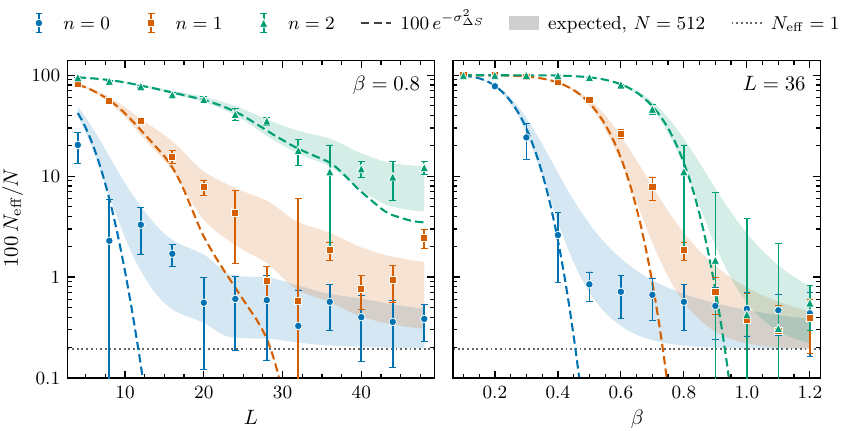}
  \caption{ The reweighting ESS, $100N_{\rm eff}/N$, of the approximate trivializing maps, from the same runs as Fig.~\ref{fig:test1} ($N=512$).
  \textbf{Left:} as a function of lattice size $L$ at fixed $\beta=0.8$. 
  \textbf{Right:} as a function of $\beta$ at fixed lattice size $L=36$. 
  Symbols: flow action truncated at order $n=0$, 1 and 2, with jackknife errors.
  Dashed curves: $100\,e^{-\sigma_{\Delta S}^2}$, Eq.~\eqref{eq:ess_gauss}.
  Bands: central 68\% range of the estimator~\eqref{eq:ess} for batches of $N=512$ Gaussian $\Delta S$.
  Both use the measured $\sigma_{\Delta S}$, interpolated monotonically between the simulated values of $L$ or $\beta$.
  Dotted line: $N_{\rm eff}=1$.
  \label{fig:ess}}
\end{figure}

We plot \(100N_{\mathrm{eff}}/N\) as a percentage; values near \(100\%\) indicate excellent overlap (low weight variance), whereas values near \(0\%\) signal weight degeneracy and unreliable estimates without far larger batches \cite{Kish1965,Liu2001,DoucetJohansen2011,ElviraMartinoRobert2022}. As seen in Fig.~\ref{fig:ess} (right), at fixed \(L=36\) the ESS decreases as \(\beta\) grows, but each higher order of the flow action substantially delays this drop, widening the coupling range with good overlap. At fixed \(\beta=0.8\), Fig.~\ref{fig:ess} (left) shows that the ESS falls rapidly with \(L\). The higher-order flows maintain usable ESS to larger \(L\), but all orders show the volume penalty expected for reweighting alone.

The curves in Fig.~\ref{fig:ess} have no free parameters: they follow from the measured $\Err$ of Fig.~\ref{fig:test1}. For large $N$, Eq.~\eqref{eq:ess} tends to $\langle w\rangle^2/\langle w^2\rangle$, with $w\propto e^{-\Delta S}$. For Gaussian $\Delta S$ with mean $\langle\Delta S\rangle$ and variance $\Err^2$, $\langle e^{-a\Delta S}\rangle=e^{-a\langle\Delta S\rangle+a^2\Err^2/2}$, so that
\begin{align}
    \frac{N_{\rm eff}}{N}\;\to\;\frac{\langle w\rangle^2}{\langle w^2\rangle}=e^{-\Err^2}
    \label{eq:ess_gauss}
\end{align}
independently of $\langle\Delta S\rangle$ (dashed curves); since $\Err\propto L$, it decays as a Gaussian in $L$. In general $\ln(\langle w\rangle^2/\langle w^2\rangle)=-\kappa_2+\kappa_3-\tfrac{7}{12}\kappa_4+\dots$ in terms of the cumulants $\kappa_n$ of $\Delta S$, with $\kappa_2=\Err^2$. For a local map all $\kappa_n$ are proportional to $|\Lambda|$, so the large-$N$ ESS falls exponentially with the volume, and the Gaussian form applies when the mismatch per site is small. For a finite batch the estimator~\eqref{eq:ess} is biased upward: $\langle w^2\rangle$ is dominated by configurations with $\Delta S\approx\langle\Delta S\rangle-2\Err^2$, which a batch of $N$ configurations rarely contains once $e^{-\Err^2}\lesssim N^{-1/2}$ ($4.4\%$ for $N=512$). Below this level the measured ESS lies above Eq.~\eqref{eq:ess_gauss} and approaches the minimum $N_{\rm eff}=1$. The bands in Fig.~\ref{fig:ess} show the central 68\% range of Eq.~\eqref{eq:ess} for batches of $N=512$ Gaussian $\Delta S$ with the measured $\Err$. The data lie within twice the band half-width at 67 of the 72 points; the largest deviation, at $n=0$ and $L=4$, is an ESS below the Gaussian expectation and indicates a non-Gaussian distribution of $\Delta S$. Where $e^{-\Err^2}\lesssim N^{-1/2}$ the measured ESS is therefore limited by the batch size and does not measure the decay of the overlap; that decay is measured directly by $\Err$ (Fig.~\ref{fig:test1}). These results show that reaching the critical region (large correlation length) on large lattices requires a substantially more accurate approximation to the trivializing flow.  

\FloatBarrier
\subsection{Transformed HMC}
\label{sec:num_thmc}

A common measure of Monte Carlo efficiency is the dynamical critical exponent $z$,
defined by the scaling of the integrated autocorrelation time with the
correlation length,
\begin{align}
    \tau_{\mathrm{int}} = \xi^z F(\xi/L),
\end{align}
where $F(\xi/L)$ is a smooth finite-size scaling function.  At fixed $\xi/L$,
the exponent $z$ characterizes critical slowing down.  Direct HMC for the
magnetic susceptibility in the two-dimensional $\mathrm{O}(3)$ model shows the expected power-law growth.  The reweighting
results above show that using an approximate trivializing map by itself does not
remove this problem: once the map lies outside its overlap window, the ESS
collapses rapidly.

We therefore also tested the second-order ($n=2$) flow action inside a
transformed-HMC algorithm with an exact accept/reject step.  In this algorithm
the Markov chain evolves the transformed variables, and the target distribution
is enforced by the transformed Hamiltonian and a Metropolis accept/reject step.
The endpoint action was computed with \texttt{gl4}, an implicit fourth-order
Runge--Kutta--Munthe-Kaas flow integrator built on the two-stage
Gauss--Legendre tableau~\cite{MUNTHEKAAS1999115,HairerLubichWanner2006}, with
$16$ steps of size $h=1/16$; the molecular-dynamics force used $8$ steps of the explicit
fourth-order Runge--Kutta--Munthe-Kaas integrator.  The
log-Jacobian is accumulated as the Gauss--Legendre quadrature of
$\sum_{x,a}[\partial^a_x]^2 S_f$ along the discrete flow,
cf.\ Eq.~\eqref{eq:jacobianODE}, and differs from the log-determinant of the
discrete map by $\mathcal{O}(h^4)$.  The chain is therefore exact for the target
defined by the discretized flow, which differs from the $\mathrm{O}(3)$
distribution by this $\mathcal{O}(h^4)$ error.  Reversibility of the flow
integrator is not required: the molecular-dynamics step is reversible and
volume preserving for any force that depends only on the position, so the
coarser flow used for the force affects only the acceptance rate.  The HMC molecular dynamics
used the second-order minimum-norm integrator, with the same integrator used for
the matched direct-HMC runs.  The transformed molecular-dynamics force was
computed from the continuum adjoint-sensitivity equations derived in
Sec.~\ref{sec:adjoint}.  The implemented basis provides analytic expressions
for the Lie gradient, Lie Laplacian, Lie gradient of the Lie Laplacian, and the
Hessian-vector products needed by the adjoint equation.  These expressions agree with automatic differentiation to a relative precision
better than $10^{-13}$.  The adjoint equation~\eqref{eq:adjointflow2} is integrated
backward with the classical fourth-order Runge--Kutta scheme, evaluated on the
stage configurations of the forward Runge--Kutta--Munthe-Kaas steps.  We compared
the resulting force and parameter gradient with the exact gradients of the
discretized transformed action and loss, obtained by reverse-mode automatic
differentiation through the same $n$-step flow map, in double precision at
$L=4$, $6$, and $8$ along $\xi/L\simeq1/4$: for the force on thermalized
configurations, and for the parameter gradient on samples of the trivial
distribution with the flow-action coefficients randomly perturbed about their
perturbative values.  The relative difference lies between $2.2\times10^{-5}$ and $2.8\times10^{-5}$
for the force at $n=8$ and between $3\times10^{-6}$ and $4\times10^{-5}$ for the
parameter gradient at $n=16$, and the force converges to that of a $256$-step
reference map as $h^4$, at the same rate as the exact gradient of the discretized action.

The direct and transformed-HMC wall-clock measurements reported below were performed with
the JAX implementation using XLA just-in-time compilation on NVIDIA A100 GPUs on
Perlmutter at the National Energy Research Scientific Computing Center (NERSC).
For these comparisons, both direct and transformed HMC used trajectories of
unit length and the second-order minimum-norm molecular-dynamics
integrator~\cite{Omelyan:2002qkh,Takaishi:2005tz}.  The trajectory length was not tuned, so
the autocorrelation reductions quoted below are relative to this fixed-length baseline.
For each lattice size we discarded 500 warmup trajectories and then measured
5000 trajectories with no thinning.  Throughout, the integrated
autocorrelation time is $\tau_{\rm int}=\tfrac12+\sum_{t\ge1}\rho(t)$, where
$\rho(t)$ is the normalized autocorrelation function, so that $N$ measurements
contain $N/(2\tau_{\rm int})$ effectively independent
samples~\cite{Madras:1988ei,Wolff:2003sm}.  The matched direct-HMC runs were performed with
128 independent streams, and their $\tau_{\rm int}(\chi_m)$ was estimated with the
$\Gamma$-method of Ref.~\cite{Wolff:2003sm}, combining the streams as replicas.  For the
transformed-HMC runs, $\tau_{\rm int}(\chi_m)$ was estimated from the autocorrelation
function of the magnetic-susceptibility history, summed up to the self-consistent
window $W\ge5\tau_{\rm int}$ of Ref.~\cite{Madras:1988ei}.  These runs used one stream because each trajectory is much more expensive; with measured
$\tau_{\rm int}(\chi_m)$ in the range $1$--$8$ trajectories, a 5000-trajectory
run yields $330$--$2150$ effectively independent samples, which is adequate for
the $\Gamma$-method estimator.

\begin{figure}[t]
\centering
  \includegraphics[width=.68\linewidth]{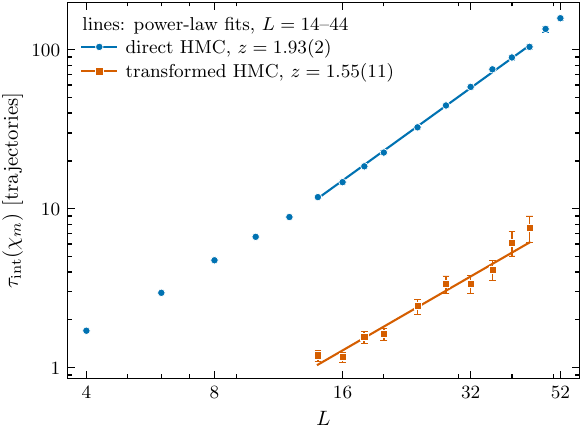}
  \caption{Integrated autocorrelation time
  $\tau_{\rm int}(\chi_m)=\tfrac12+\sum_{t\ge1}\rho(t)$ in trajectory units along the
  approximately fixed $\xi/L\simeq 1/4$ line.
  Circles: direct HMC ($L=4$--$52$; independent runs of a PyTorch implementation
  with 64 streams of $10^4$--$10^5$ unthinned trajectories and $N_{\rm md}=3$,
  analyzed with the $\Gamma$-method of Ref.~\cite{Wolff:2003sm} combined over
  streams).  Squares: transformed HMC ($L=14$--$44$; single chains of 5000
  trajectories, MD step count chosen to keep the Metropolis
  acceptance high, $0.92$--$0.98$; error bars $\tau_{\rm int}\sqrt{(4W+2)/N}$ with
  $W=6\tau_{\rm int}$).  Lines: weighted power-law fits over $L=14$--$44$,
  $z=1.93(2)$ (direct) and $z=1.55(11)$ (transformed).}
  \label{fig:thmc_iat}
\end{figure}

The transformed-HMC runs were performed along a line of approximately fixed
$\xi/L=1/4$.  The MD step count was chosen in coarse steps to keep the acceptance
high; for $L=14,\ldots,44$ the final tuned runs have acceptances between $0.92$
and $0.98$.  As shown in Fig.~\ref{fig:thmc_iat}, the transformed chain
substantially reduces the autocorrelation time of the magnetic susceptibility
when measured in trajectory units.  Reversibility violations are at the
level of floating-point roundoff.  Weighted power-law fits to both datasets of Fig.~\ref{fig:thmc_iat} over the
common range $L=14$--$44$ give
\begin{align}
    \tau_{\mathrm{int}}^{\mathrm{direct}}(\chi_m) &\sim L^{1.93(2)}, \\
    \tau_{\mathrm{int}}^{\mathrm{tr}}(\chi_m) &\sim L^{1.55(11)} ,
\end{align}
with $\chi^2/\mathrm{dof}=1.1$ and $1.2$, respectively.  Over this range the
transformed chain has $\tau_{\rm int}(\chi_m)$ smaller by factors of
$10$--$18$, so the transformed map is effective as a decorrelating change of
variables over the range studied.  The direct-HMC data are not described by a single power
law over the full range: the local exponent increases from about $1.5$ at
$L\le12$ to about $2$ at $L\ge20$, where a fit over $L=20$--$52$ gives
$z=2.02(3)$.  The direct-HMC runs of Fig.~\ref{fig:thmc_iat} used a fixed
$N_{\rm md}=3$, whereas $N_{\rm md}^{\rm tr}$ was chosen to keep the
transformed-HMC acceptance high ($0.92$--$0.98$).  The matched direct-HMC runs of
Table~\ref{tab:matched_cost}, with $N_{\rm md}$ tuned to acceptances $0.93$--$0.99$, give
$\tau_{\rm int}^{\rm direct}(\chi_m)\sim L^{1.79(4)}$ over the same range
($\chi^2/\mathrm{dof}=0.5$), so part of the difference between the two exponents above
comes from the decreasing acceptance at fixed $N_{\rm md}$, and the change of variables
reduces the exponent by $0.24(11)$.  This reduction is relative to direct HMC with the same fixed
trajectory length.  For free fields, the integrated autocorrelation time of the magnetic
susceptibility in HMC with random trajectory lengths of fixed mean grows as $\xi^2$, so that $z=2$,
while a mean trajectory length proportional to $\xi$ makes the cost in molecular-dynamics time grow only as
$\xi$~\cite{Kennedy:2000ju}.  The direct-HMC exponent at $L\ge20$, $z=2.02(3)$, is close to this
fixed-length value.  The trajectory length was not tuned for either algorithm, so
these results do not establish a smaller exponent than that of direct HMC with a tuned
trajectory length.

The wall-clock comparison is less favorable.  To separate autocorrelation
reduction from implementation cost, we ran matched direct-HMC simulations in
which $N_{\rm md}$ was the smallest value in $\{1,2,3,4,6,8\}$ whose acceptance in a test run
of $1000$ trajectories reached that of the transformed-HMC run at the same lattice size.
Here $N_{\rm md}$ and $N_{\rm md}^{\rm tr}$ are the numbers of molecular-dynamics steps per
unit-length trajectory; the flow integrations inside each evaluation of the transformed force
or endpoint action are not counted.  Direct HMC required only
$N_{\rm md}=3,\ldots,8$ over $L=14,\ldots,44$, while transformed HMC used
$N_{\rm md}^{\rm tr}=8,\ldots,80$, chosen in coarse steps to keep the acceptance high rather
than minimized.  With $N_{\rm md}^{\rm tr}=16$, the transformed-HMC run at $L=16$ has the
highest acceptance in Table~\ref{tab:matched_cost}, $0.978$, and the matched direct-HMC run
needs $N_{\rm md}=6$ to reach it; this is why both step counts are larger at $L=16$ than
at $L=18$.  Table~\ref{tab:matched_cost} summarizes the
matched comparison.  The transformed chain reduces $\tau_{\rm int}$ by
roughly an order of magnitude, but the trajectory cost is much larger.  The
resulting cost proxy, $t_{\rm traj}\tau_{\rm int}$, is larger for
transformed HMC by factors of about $16$--$40$ in the matched runs.  Where the
matched direct-HMC runs also use $N_{\rm md}=3$ ($L=14$ and $18$), their $\tau_{\rm int}$ agree with
the longer runs of Fig.~\ref{fig:thmc_iat} within errors, and the two sets agree up to $L=28$.
At $L\ge32$ the fixed-$N_{\rm md}=3$ runs, whose acceptance decreases with $L$, give
$\tau_{\rm int}$ larger by $12$--$25\%$.

\begin{table}[t]
\centering
\begin{tabular}{rrrrrrrrrrr}
\hline
$L$ & $\beta$ & $N_{\rm md}$ & $N_{\rm md}^{\rm tr}$ & $A$ & $A^{\rm tr}$ & $\tau_{\rm int}$ & $\tau_{\rm int}^{\rm tr}$ & $t_{\rm traj}$ & $t_{\rm traj}^{\rm tr}$ & cost ratio \\
& & & & & & & & [ms] & [ms] & \\
\hline
14 & 1.234 & 3 & 8 & 0.951 & 0.921 & 11.75(31) & 1.195(94) & 7.00 & 1175 & 17.1(1.4) \\
16 & 1.273 & 6 & 16 & 0.985 & 0.978 & 14.50(42) & 1.164(90) & 7.20 & 2130 & 23.8(2.0) \\
18 & 1.303 & 3 & 12 & 0.933 & 0.922 & 18.35(59) & 1.56(14) & 7.18 & 1337 & 15.8(1.5) \\
20 & 1.331 & 4 & 16 & 0.957 & 0.937 & 22.19(77) & 1.63(15) & 7.11 & 1647 & 17.0(1.7) \\
24 & 1.375 & 4 & 20 & 0.948 & 0.932 & 31.3(1.3) & 2.43(27) & 7.47 & 1983 & 20.6(2.4) \\
28 & 1.411 & 4 & 24 & 0.939 & 0.934 & 42.0(2.0) & 3.35(43) & 7.55 & 2135 & 22.5(3.1) \\
32 & 1.4403 & 6 & 32 & 0.968 & 0.940 & 52.0(2.7) & 3.37(43) & 7.54 & 2696 & 23.2(3.2) \\
36 & 1.466 & 6 & 44 & 0.964 & 0.943 & 61.0(3.3) & 4.12(58) & 7.41 & 2911 & 26.5(4.0) \\
40 & 1.4866 & 6 & 60 & 0.959 & 0.959 & 71.6(4.2) & 6.1(1.1) & 8.04 & 3464 & 36.7(6.7) \\
44 & 1.5054 & 8 & 80 & 0.975 & 0.961 & 92.8(6.1) & 7.6(1.4) & 8.39 & 4134 & 40.2(8.1) \\
\hline
\end{tabular}

\caption{Matched direct-HMC and transformed-HMC comparison.  Both algorithms use
the second-order minimum-norm MD integrator.  $N_{\rm md}$, $A$ and $\tau_{\rm int}$ are the MD
step count, the Metropolis acceptance and $\tau_{\rm int}(\chi_m)$ in trajectory units for direct
HMC; the superscript tr denotes transformed HMC.  $t_{\rm traj}$ is the wall-clock time of one
trajectory of all 128 direct-HMC streams and $t_{\rm traj}^{\rm tr}$ that of one single-stream
transformed trajectory, both including the measurement of $\chi_m$ after every trajectory
(per stream the direct cost is a further factor of 128 smaller).
The last column is the ratio of $t_{\rm traj}\tau_{\rm int}$ for transformed HMC to the matched
direct-HMC value.  Errors on $\tau_{\rm int}$ are from the $\Gamma$-method with the 128 streams
combined as replicas~\cite{Wolff:2003sm}; errors on $\tau_{\rm int}^{\rm tr}$ are
$\tau_{\rm int}^{\rm tr}\sqrt{(4W+2)/N}$ with $W=6\tau_{\rm int}^{\rm tr}$ and $N=5000$, as in
Fig.~\ref{fig:thmc_iat}.  The cost-ratio errors combine the two.}
\label{tab:matched_cost}
\end{table}

Weighted fits of the measured cost proxy over $L=14,\ldots,44$, with the errors of
Table~\ref{tab:matched_cost}, give
\begin{align}
    z_{\rm cost}^{\rm direct}=1.90(4),\qquad
    z_{\rm cost}^{\rm tr}=2.38(18),
\end{align}
with $\chi^2/{\rm dof}=0.7$ and $3.4$; the second error is scaled by
$\sqrt{\chi^2/{\rm dof}}$.

These wall-clock times measure latency more than arithmetic.  For direct HMC, one
trajectory of all 128 streams takes 0.3--1.0\,ms of the measured 7.0--8.4\,ms; the
rest is the measurement after every trajectory, and the single transformed-HMC stream leaves
the GPU largely idle.  The exponents $z_{\rm cost}$ above therefore characterize the present
implementation on this hardware rather than the algorithms.  To remove latency, we timed both
codes with the settings of Table~\ref{tab:matched_cost}, running many independent streams in one batch
on the same A100, and took the cost of one additional stream-trajectory, without measurements.
It is 0.4--8.6\,$\mu$s for direct HMC in single precision, as in the runs of
Table~\ref{tab:matched_cost} (0.7--17\,$\mu$s in double precision), and 43--1900\,ms for
transformed HMC, whose per-stream cost is lowest at 64 streams for $L\le32$ and at 8 streams
for $L\ge36$; larger batches run out of memory or slow down.  Per independent sample,
transformed HMC is then more expensive by factors of $3.9\times10^{3}$--$2.2\times10^{4}$ ($1.8\times10^{3}$--$1\times10^{4}$
against direct HMC in double precision).  A count of the floating-point operations in the
compiled kernels gives the hardware-independent factor $4.4\times10^{3}$--$1.4\times10^{4}$: one transformed
force costs as much as about $3.9\times10^{4}$ direct-HMC forces, and one evaluation of the endpoint
action as about $8.6$ transformed forces.  A cheaper endpoint-action integrator does not
change this.  The explicit fourth-order integrator with $64$ steps evaluates the endpoint action
with a smaller discretization error than \texttt{gl4} with $16$ steps, at 15--17\% of its
CPU time and 28\% of its floating-point operations, but lowers the operation count per
transformed trajectory and measurement by only
7--37\%, because the transformed force dominates.

Thus, the data do not support a favorable large-volume wall-clock crossover for the
present implementation.  The analytic trivializing map reduces autocorrelations
substantially in trajectory units, but the cost of transformed-force and
endpoint-action evaluations dominates the wall-clock time.  For the present
implementation, this cost outweighs the decorrelation gain.

\FloatBarrier
\subsection{Nonperturbative tuning}
\label{sec:num_np}

We explored nonperturbative tuning of the second-order flow action.  We write it as
$S_f[s,t]=\sum_k a_k(t)\,\Phi_k[s]$, a sum over the twelve functions $\Phi_k$ of the second-order flow action of
Sec.~\ref{sec:trivialFlow}, and keep the small-$t$ form of the analytic solution.  A function that first appears at order
$t^{n_k}$, $n_k=1,2$, has $a_k(t)=\bar c_k\,t^{n_k}+t^{n_k+2}P_k(t^2)$, and $\Psi\e{0}$, the only function present at
$t=0$, has $a(t)=\bar c+\bar c\,'t^2+t^4P(t^2)$; each $P_k$ is a polynomial of degree four in $t^2$.  The thirteen
coefficients $\bar c$ start from their analytic values and the sixty tail coefficients from zero, so that training starts
from the analytic map.  We minimized $\mathrm{Var}(\Delta S)$ with the Adam optimizer, using a new batch of $32$--$512$
trivially distributed configurations at every step.  The gradients are exact derivatives of the discretized flow:
$\Delta S$ is computed with $16$ steps ($h=1/16$) of the fourth-order Runge--Kutta--Munthe-Kaas integrator and
differentiated by reverse-mode automatic differentiation through these steps, i.e., with the discrete adjoint of the
integrator.  At the final parameters of six of the runs ($\beta=0.8$ and the $\xi/L\simeq1/4$ line, each at $L=4$, $8$ and $16$),
these gradients agree with those of the continuous adjoint equations of Sec.~\ref{sec:adjoint} to a relative
difference of $2.5\times10^{-6}$--$1.1\times10^{-4}$.  Retraining seven of the runs ($\beta=0.8$ at $L=4$, $8$ and $16$,
and the $\xi/L\simeq1/4$ line at $L=4$, $6$, $8$ and $14$) with the continuous adjoint instead, with all other settings
unchanged, gives tuned maps whose $\sigma_{\Delta S}$, measured on the independent batch described below, agrees with that
of the discrete-adjoint maps to within $0.2\%$.  On one A100 GPU, at $\beta=0.8$ and $L=8$ ($16$), the continuous adjoint
needs $0.61$ ($0.57$) of the time per training step of the discrete adjoint, $0.38$ ($0.53$) of its compilation time,
and about half of its host memory.  The variance is the leading term of the KL divergence,
$D_{\rm KL}[q\|p]=\langle\Delta S\rangle+\ln\langle e^{-\Delta S}\rangle=\tfrac12\mathrm{Var}(\Delta S)-\tfrac16\kappa_3
+\tfrac1{24}\kappa_4-\cdots$, where $\kappa_n$ are the cumulants of $\Delta S$ over trivially distributed $\sigma$; both
vanish only for a perfect flow, and their minima differ in general.  Each run first trained the sixty tail coefficients for
$1000$ steps with the $\bar c$ fixed; all runs except one ($L=14$ on the $\xi/L\simeq1/4$ line) then trained all $73$
coefficients for $3000$ further steps, starting from the best first-stage parameters.  The tuned map is the parameter set
with the smallest $\sigma_{\Delta S}$ on a fixed evaluation batch, checked every $25$ steps.  Because this selection favors
the fluctuations of that batch, all results below were measured on an independent batch of $N=262144$ trivially
distributed configurations, for the analytic and the tuned map alike, with the integrator settings of
Figs.~\ref{fig:test1} and~\ref{fig:ess}.

\begin{figure}[htbp]
\centering
  \includegraphics[width=\linewidth]{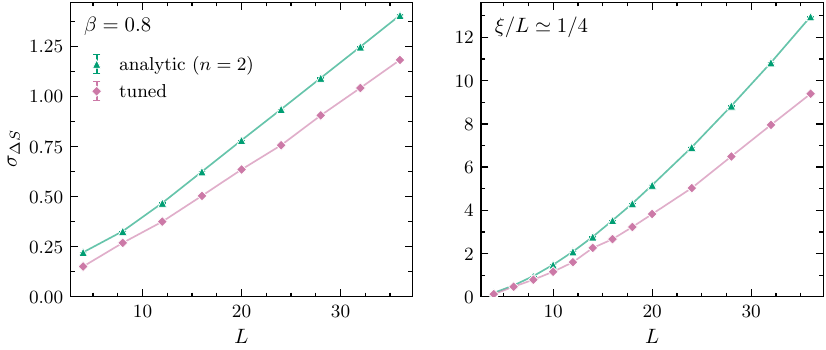}
  \caption{Standard deviation $\sigma_{\Delta S}$ of the analytic second-order map (triangles) and of the tuned map
  (diamonds), both measured on the same independent batch of $N=262144$ trivially distributed configurations; the
  jackknife errors are smaller than the symbols, and lines connect the points.
  \textbf{Left:} $L$ scan at fixed $\beta=0.8$.
  \textbf{Right:} $L$ scan along the $\xi/L \simeq 1/4$ line ($\beta$ increases with $L$).
  Tuning reduces $\sigma_{\Delta S}$ by $16$--$20\%$ at $\beta=0.8$ for $L\ge8$ and by $12$--$28\%$ along the
  $\xi/L\simeq1/4$ line for $L\ge6$ ($32\%$ and $33\%$, respectively, at $L=4$).
  \label{fig:np_tuned_std}}
\end{figure}

\begin{figure}[htbp]
\centering
  \includegraphics[width=\linewidth]{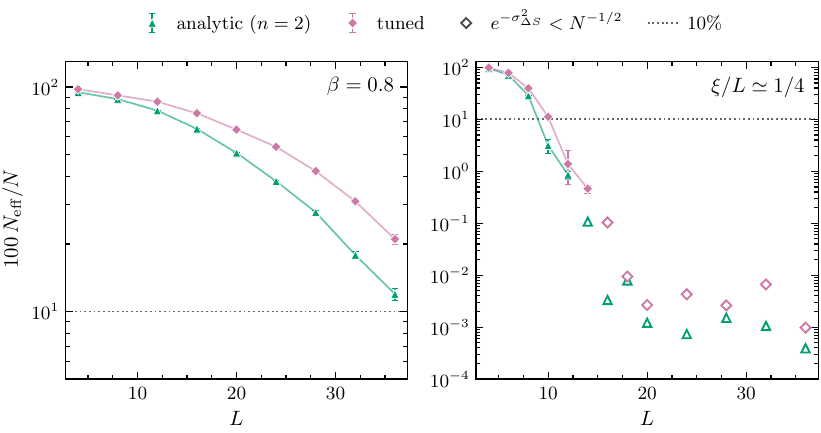}
  \caption{Reweighting ESS, $100\,N_{\rm eff}/N$, of the analytic second-order map (triangles) and of the tuned map
  (diamonds), on the batch of Fig.~\ref{fig:np_tuned_std}, with jackknife errors.
  \textbf{Left:} $L$ scan at fixed $\beta=0.8$.
  \textbf{Right:} $L$ scan along the $\xi/L \simeq 1/4$ line.
  The dotted horizontal line marks $10\%$ ESS.  Open symbols mark points with $e^{-\sigma_{\Delta S}^2}<N^{-1/2}$,
  whose ESS is set by the finite batch rather than by the overlap (see the discussion below Eq.~\eqref{eq:ess_gauss});
  they include all points with fewer than $10$ effective samples.  Lines connect the filled points.
  Tuning improves the ESS modestly at $\beta=0.8$, but the tuned ESS remains
  below $1\%$ for $L \ge 14$ along the $\xi/L \simeq 1/4$ line.
  \label{fig:np_tuned_ess}}
\end{figure}

Figures~\ref{fig:np_tuned_std} and~\ref{fig:np_tuned_ess} show $\sigma_{\Delta S}$
and the reweighting ESS of the analytic and the tuned map.  At fixed $\beta=0.8$ (left
panels), tuning reduces $\sigma_{\Delta S}$ by $16$--$20\%$ for $L\ge 8$ ($32\%$ at $L=4$), with the smallest
reduction at the largest lattice ($16\%$ at $L=36$).  For $L\ge12$, $\sigma_{\Delta S}$ grows linearly with $L$ for
both maps, with a slope that tuning lowers by $14\%$.  The ESS rises correspondingly, from $51\%$ to $65\%$ at $L=20$ and from $18\%$ to $31\%$ at $L=32$;
at this coupling the ESS exceeds $10\%$ at every $L\le36$ before and after tuning ($12\%$ and $21\%$ at $L=36$).

Along the $\xi/L \simeq 1/4$ line (right panels) $\beta$ increases with $L$, and $\sigma_{\Delta S}$ grows faster
than linearly.  Tuning reduces $\sigma_{\Delta S}$ by $12$--$28\%$ for $L\ge6$ ($33\%$ at $L=4$), by more at larger $L$;
the reduction at $L=14$ ($19\%$), the one run trained in a single stage, is smaller than at its neighbors.  The ESS of the
analytic map falls below $10\%$ at $L=10$ [$3.1(1.0)\%$], and tuning raises it to $11.2(5)\%$.  At $L=12$ the ESS is about
$1\%$ for both maps [$0.86(16)\%$ and $1.4(1.1)\%$], and for $L\ge14$ the tuned ESS is below $1\%$.  From $L=14$ for
the analytic map and $L=16$ for the tuned map, $e^{-\sigma_{\Delta S}^2}<N^{-1/2}$, so that the measured ESS is set by the
finite batch rather than by the overlap (see the discussion below Eq.~\eqref{eq:ess_gauss}); these points are drawn as
open symbols in Fig.~\ref{fig:np_tuned_ess}.  Nonperturbative training of the coefficients therefore provides a genuine
but incremental improvement---reducing $\sigma_{\Delta S}$ by $16$--$20\%$ at $\beta=0.8$ and by $12$--$28\%$ along
$\xi/L\simeq1/4$ (for $L\ge 8$ and $L\ge 6$, respectively), and extending the useful volume window (ESS $>10\%$) along
$\xi/L\simeq1/4$ by one step in $L$, from $L\le8$ to $L\le10$---but $\sigma_{\Delta S}$ still grows with $L$, and the
reweighting ESS still decays exponentially with the volume.

\FloatBarrier
\subsection{Approximate effective action}
\label{sec:num_sur}

We now turn to the approximate effective action required by the algorithm of Sec.~\ref{sec:EffectiveAction}, which we call the surrogate effective action, $S_{\rm sur}\equiv{\cal S}_0^{\rm ap}\approx S_{\rm eff}\equiv{\cal S}_0[\sigma,1]$. We constructed surrogates for the second-order flow action along the
  $\xi/L \simeq 1/4$ line for $L = 8,\ldots,28$. The truncated Galerkin coefficients of Sec.~\ref{sec:EffectiveAction} are not used: the effective action of the second-order flow action starts at ${\cal O}(\beta^4)$, the twelve-function system reproduces only the projection of this leading term, and its higher orders depend on the truncation. Instead, the coefficients are determined nonperturbatively, and two architectures were studied.
  A \emph{linear surrogate} expands $S_{\rm sur}$ in the twelve functions of Sec.~\ref{sec:trivialFlow} and four two-spin shells $\sum_x\sum_{\delta}s\tran_x s_{x+\delta}$, where $\delta$ runs over the lattice vectors obtained from $(2,1)$, $(2,2)$, $(3,1)$ or $(3,2)$ by reflections and exchange of the axes; like $\Psi\e{0}$, the shells are eigenfunctions of the Lie Laplacian. Its coefficients minimize the KL divergence of Eq.~\eqref{eq:KLsur}, estimated on configurations generated by HMC with $S_{\rm sur}$, so that ${\rm Cov}(\Psi\e{[k]},\Delta S)=0$ for every function of the expansion [Eq.~\eqref{eq:KLgrad}], with $\Delta S = S_{\rm eff} - S_{\rm sur}$. The minimum is found iteratively: each iteration generates new configurations with the current coefficients and takes a Gauss--Newton step, which uses the covariance matrix of the $\Psi\e{[k]}$ in place of the full Hessian of Sec.~\ref{sec:adjoint}.  A \emph{CNN surrogate} applies a periodic convolutional network to the rotation-invariant products $s\tran_x s_{x+\delta}$,
  averaged over each of six to nine shells of displacements $\delta$, and sums its output over the lattice; it has three to five
  hidden layers of 32 channels with $3\times3$ kernels whose dilations grow with $L$. Its parameters minimize ${\rm Var}(\Delta S)$
  over trivially (Haar) distributed $\sigma$, i.e.\ the CNN surrogate is a least-squares fit to $S_{\rm eff}$ under the Haar
  measure rather than a minimum of the KL divergence of Eq.~\eqref{eq:KLsur}.

\begin{figure}[htbp]
  \centering
    \includegraphics[width=.495\linewidth]{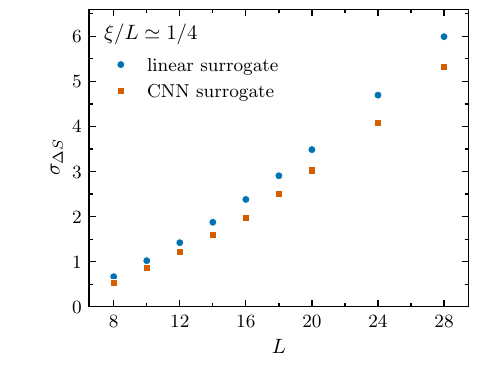}
    \includegraphics[width=.495\linewidth]{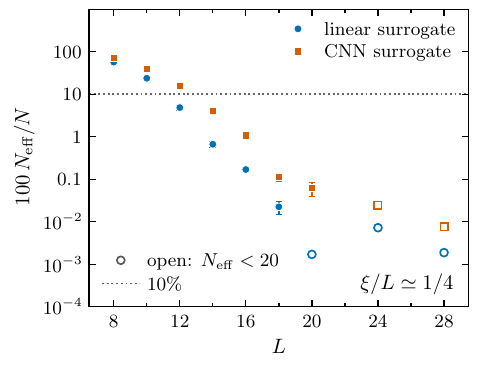}
    \caption{Quality of the nonperturbative surrogate effective action at fixed
    $\xi/L \simeq 1/4$.
    \textbf{Left:} Standard deviation $\sigma_{\Delta S}$ of the residual
    $\Delta S = S_{\rm eff} - S_{\rm sur}$ vs.\ $L$.  Error bars, $\sigma_{\Delta S}/\sqrt{2(N-1)}$, are the standard error for $N$ independent Gaussian samples
    and are smaller than the symbols.  Successive surrogate-HMC measurements are nearly independent: on all points whose
    measurement histories were stored (all except the linear surrogate at $L=16$, 20, 24 and 28),
    the integrated autocorrelation time of $(\Delta S-\langle\Delta S\rangle)^2$, which controls the error of $\sigma_{\Delta S}$, is
    $0.50$--$0.53$ trajectories ($1/2$ for independent measurements), and the $\Gamma$-method~\cite{Wolff:2003sm} with the
    chains as replicas, which also accounts for the non-Gaussian tails of $\Delta S$, enlarges the errors by at most $7\%$.
    \textbf{Right:} Reweighting ESS vs.\ $L$ (logarithmic scale), with jackknife errors from 100 blocks of consecutive
    measurements (no error bar for the linear surrogate at $L=16$, whose history was not stored); the $10\%$ reference is shown dotted.  Filled symbols use $N = 6.4 \times 10^{5}$--$1.28 \times 10^{7}$
    configurations.  Open symbols are lower-statistics estimates ($N = 2.56 \times 10^{5}$ linear, $6.4 \times 10^{4}$ CNN)
    that rest on fewer than 20 effective samples and indicate only that the ESS is below about $0.03\%$.  The CNN
    surrogate consistently achieves smaller $\sigma_{\Delta S}$ and higher ESS
    than the linear surrogate, but both architectures lose usable overlap for
    $L \gtrsim 14$.
    \label{fig:surrogate_quality}}
  \end{figure}

  The surrogate quality is measured by $\sigma_{\Delta S}$ and by the reweighting
  ESS (Eq.~\eqref{eq:ess}), the two quantities that determine the efficiency of the algorithm of Sec.~\ref{sec:EffectiveAction}.  Figure~\ref{fig:surrogate_quality} shows both
  quantities vs.\ $L$.  Compared with the flow alone ($S_{\rm sur}=0$ with Haar-distributed $\sigma$, for which $\sigma_{\Delta S}=0.96$ at $L=8$ and $3.54$ at $L=16$), the linear surrogate lowers $\sigma_{\Delta S}$ by $30$--$33\%$ at every $L$ and raises the ESS from $29\%$ to $57\%$ at $L=8$, from $3\%$ to $24\%$ at $L=10$ and from $0.9\%$ to $4.9\%$ at $L=12$, so that the window $\mathrm{ESS}>10\%$ grows from $L\le8$ to $L\le10$.  The CNN surrogate reduces  $\sigma_{\Delta S}$ by roughly
  $10$--$20\%$ compared with the linear surrogate (larger reductions at smaller $L$),
  and correspondingly extends the usable volume window ($\mathrm{ESS} > 10\%$) by
  approximately one step in $L$: the CNN achieves $\mathrm{ESS} \approx 72\%$ at
  $L = 8$ and $\mathrm{ESS} \approx 16\%$ at $L = 12$, compared with $\approx 57\%$
  and $5\%$ for the linear surrogate.  For both architectures the ESS collapses to
  $\lesssim 1\%$ for $L \geq 16$, reflecting the same approximately linear growth
  of $\sigma_{\Delta S}$ with $L$ observed for the perturbative maps.  For the linear surrogate, $\sigma_{\Delta S}$ rises from $0.67$ at $L=8$ to $2.38$ at $L=16$; by the estimate of Sec.~\ref{sec:EffectiveAction}, the exact algorithm would then accept about $64\%$ and $9\%$ of proposals that decorrelate $\Delta S$.

\FloatBarrier
\subsection{Multi-timestep HMC}
\label{sec:num_mts}

\begin{figure}[t]
  \centering
    \includegraphics[width=.68\linewidth]{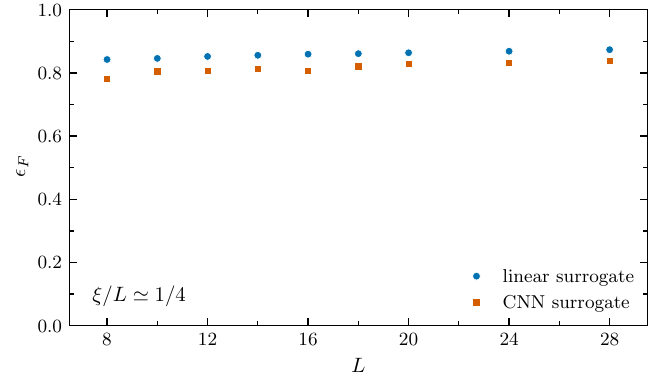}
    \caption{Relative RMS force error $\epsilon_F$ vs.\ $L$ at fixed
    $\xi/L \simeq 1/4$ for both surrogates.  For all volumes and both
    architectures $\epsilon_F \approx 0.78$--$0.87$, showing that neither
    surrogate provides a useful force approximation for multi-timestepping HMC.  Statistical errors are smaller than the symbols.
    \label{fig:surrogate_force_error}}
  \end{figure}

  A standard strategy for exploiting a cheap surrogate action in HMC is
  multi-timestepping (MTS-HMC): inner MD steps use the inexpensive force
  $\nabla S_{\rm sur}$, while a correction step applies the residual
  $\nabla(S_{\rm eff} - S_{\rm sur})$ much less frequently.  The scheme is
  efficient only when the relative force error
  \begin{equation}
    \epsilon_F = \frac{\bigl\|\nabla S_{\rm eff} - \nabla S_{\rm sur}\bigr\|_{\rm rms}}
                      {\bigl\|\nabla S_{\rm eff}\bigr\|_{\rm rms}}
  \end{equation}
  is small.  We measured $\epsilon_F$ by running surrogate HMC and evaluating both
  forces on the generated trajectories; the results are shown in
  Fig.~\ref{fig:surrogate_force_error}.  For both architectures $\epsilon_F \approx
  0.78$--$0.87$ at all volumes studied; the residual force is therefore
  $78$--$87\%$ of the full effective-action force in RMS norm.  This large residual is not surprising:
  both surrogates are fitted to a scalar objective, which imposes no
  direct constraint on the $2L^2$ independent force components.  MTS-HMC inner
  steps would require $\epsilon_F \ll 1$, which neither surrogate achieves.

\FloatBarrier
\section{Conclusions}
\label{sec:conclusions}

In this paper, we developed and tested a framework for constructing trivializing maps for the two-dimensional $\mathrm{O}(3)$ model based on gradient flow. On the analytic side, we expanded the flow action to second order in flow time, organized by eigenfunctions of the Lie Laplacian, and derived compact expressions for the Jacobian and for the time evolution of the effective action induced by the flow. We also derived and implemented the adjoint-sensitivity equations needed for transformed-HMC forces and for variational derivatives of the flow-action coefficients. Together, these pieces provide a systematic route to computable flows and their effective actions.

Numerically, we quantified the accuracy of truncated flows by the standard deviation of the action mismatch, $\sigma_{\Delta S}$. At fixed coupling, $\sigma_{\Delta S}$ grows approximately linearly with lattice size $L$; at fixed volume, it grows polynomially with $\beta$, while higher-order truncations reduce the overall error. Reweighting diagnostics show the expected deterioration of ensemble overlap with increasing $L$ and $\beta$, again mitigated, but not eliminated, by including higher orders. Approximate trivializing maps therefore have a useful finite window, but standalone reweighting is not a viable asymptotic algorithm.

Embedding the same map in an exact transformed-HMC algorithm changes the diagnostic. The transformed chain is exact up to the $\mathcal{O}(h^4)$ discretization error of the flow integration, has high acceptance after increasing the number of MD integration steps, and substantially reduces the integrated autocorrelation time of the magnetic susceptibility in trajectory units over the tested range, relative to direct HMC with the same fixed trajectory length. This demonstrates that the flow is useful as a decorrelating change of variables inside an exact MCMC algorithm. However, the wall-clock cost is dominated by transformed-force and endpoint-action evaluations. In matched-acceptance comparisons with direct HMC, the present transformed-HMC implementation is slower by a factor of 16--40 in the measured, latency-bound wall-clock time and by three to four orders of magnitude per independent sample at full GPU throughput or by operation count; it also has a larger fitted cost exponent over the measured range.

We also tested two natural improvements suggested by the formalism: nonperturbative tuning of the flow-action coefficients with adjoint gradients, and surrogate effective actions for cheaper transformed-HMC forces. In the basis used here, neither produced a competitive algorithm. Adjoint tuning of the flow-action coefficients reduced $\sigma_{\Delta S}$ by $12$--$28\%$ for $L\ge6$ and extended the usable volume window at fixed $\xi/L$ by one step in $L$, but $\sigma_{\Delta S}$ still grows with $L$ within the fixed perturbative operator basis; the exponential decay of the reweighting ESS with volume was not overcome. Surrogate effective actions (linear basis and CNN) modestly extended the usable overlap window, but their forces had relative RMS error $\epsilon_F \approx 0.78$--$0.87$ at all volumes studied---far above the $\epsilon_F \ll 1$ required for multi-timestepping HMC---because their training objectives, a Kullback--Leibler divergence for the linear basis and $\mathrm{Var}(\Delta S)$ for the CNN, are scalar and impose no direct constraint on force accuracy. These negative results are useful: they identify the force/action cost and the limited expressiveness of the operator basis as the main bottlenecks, rather than the correctness of the analytic construction.

The conclusion is therefore conservative. The analytic trivializing-map framework works, and the adjoint formalism can be implemented and validated. Approximate maps can reduce autocorrelations when embedded in exact HMC, but the present $\mathrm{O}(3)$ implementation does not beat direct HMC in wall-clock time. Future progress will require either a more accurate symmetry-adapted operator basis, a genuinely cheaper transformed-force computation, or a multiscale construction that extends the useful overlap window without increasing the force cost prohibitively. The $\mathrm{O}(3)$ sigma model remains a useful testing ground for these ideas before they are applied to gauge theories.

\begin{acknowledgments}

   This work is supported in part by U.S. DOE grant \mbox{
  DE-FG02-04ER41302}.
      This work was performed in part using computing
facilities at William \& Mary which were provided by
contributions from the National Science Foundation (MRI grant
PHY-1626177), the Commonwealth of Virginia Equipment Trust Fund and
the Office of Naval Research.
The authors acknowledge William \& Mary Research Computing for providing computational resources and/or technical support that have contributed to the results reported within this paper.
This research used resources of the National Energy Research Scientific Computing Center (NERSC) through the ERCAP project ``{\tt hadron}''. NERSC is a U.S. Department of Energy Office of Science User Facility located at Lawrence Berkeley National Laboratory, operated under Contract No.~DE-AC02-05CH11231.

\end{acknowledgments}

\appendix*
\section{\texorpdfstring{$\mathrm{SO}(3)$}{SO(3)} flows and Jacobians}

\label{sec:SO3}
In this appendix, we collect useful identities for $\mathrm{SO}(3)$ and derive the expression for the Jacobian matrix that appears in the change 
of variables induced by the gradient flow.

\subsection*{\texorpdfstring{$\mathrm{SO}(3)$}{SO(3)} generators}
The $\mathrm{SO}(3)$ generators are $3\times 3$ traceless antisymmetric matrices $L_a$ that are normalized as
\begin{align}
    \Tr(L_a L_b) = - 2 \delta_{ab}
\end{align}
The $\mathfrak{so}(3)$  algebra is
\begin{align}
    [L_a,L_b] &= \epsilon_{abc} L_c
    \label{eq:so3LieAlgebra}
\end{align}
The structure constants  $\epsilon_{abc}$ are the components of the totally antisymmetric Levi-Civita tensor.
Furthermore,
\begin{align}
    \sum_a L_a^2 = - C_F {\mathbf 1}_3 \,
    \label{eq:Casimir}
\end{align}
where $C_F=2$ is the eigenvalue $l(l+1)$, with $l=1$, of the quadratic Casimir $-\sum_a L_a^2$ in the defining (vector) representation of $\mathrm{SO}(3)$.
Finally, we have
\begin{align}
\label{eq:compSO3}
    \sum_a (L_a)_{ij}(L_a)_{kl} = \delta_{ik}\delta_{jl} - \delta_{il}\delta_{jk} 
\end{align}

\subsection*{The Jacobian}

Let $R_x \in \mathrm{SO}(3)$ denote the flow variables and 
$\Omega_x \in \mathrm{SO}(3)$ their preimages such that
\begin{align}
    R={\cal G}[\Omega]
\end{align}
This map defines a change of variables.
On each site, we define the right-invariant Maurer--Cartan one-forms as
\begin{align}
    \theta_x^a =& -\frac{1}{2} \Tr [ d R_x R_x\tran L_a ]\\
    \hat \theta_x^a =& -\frac{1}{2} \Tr [ d \Omega_x \Omega_x \tran L_a ]
\end{align}
In addition, the total differential of any function of an $\mathrm{SO}(3)$ field is  
\begin{align}
    d {\cal F}[R] = \sum_{a,x} \theta_x^a \partial^a_x {\cal F}[R]
\end{align}
From these formulas, we have 
\begin{align}
    \theta_x^a = -\frac{1}{2} \Tr [ d R_x R_x\tran L_a ]=  \sum_{b,y} \left(-\frac{1}{2} \Tr [ \hat\partial_y^b R_x R_x\tran L_a] \hat\theta_y^b\right)
\end{align}
The Jacobian matrix is the linear transformation between one-forms; therefore,
\begin{align}
{\cal J}^{a,b}(x,y) =  -\frac{1}{2} \Tr [ \hat\partial_y^b R_x R_x\tran L_a]
\label{eq:apendix_jac}
\end{align}
and
\begin{align}
\theta^a_x = \sum_{b,y}{\cal J}^{a,b}(x,y) \hat\theta^b_y
\end{align}
The corresponding transformation of derivatives is
\begin{align}
\hat \partial_x^a = \sum_{b,y} {\cal J}^{b,a}(y,x)\partial_y^b
\end{align}

\subsection*{The Jacobian of a map defined by a flow}

Consider now a time-dependent map
\begin{align}
    U_t={\cal G}_t[V]
\end{align}
where $U_t$ is the solution of the flow equation
\begin{equation}
    \dot U_x = \sum_a L_a Z^a_x U_x = Z_x U_x \,.
    \label{eq:UflowApp}
\end{equation}
In this case

\begin{align}
    \frac{d \ln\det [{\cal J}[U_t] ]}{dt} =\frac{d \Tr \ln[{\cal J}[U_t]]}{dt} = \Tr [ \dot {\cal J}[U_t] {\cal J}^{-1}[U_t]]
\end{align}
where $U_t$ is the element of $\mathrm{SO}(3)$ at flow time $t$. Differentiating the Jacobian matrix in Eq.~\eqref{eq:apendix_jac}, we obtain
\begin{align}
\dot{\cal J}_t^{a,b}[V](x,y) = - \frac{1}{2} \left\{ \Tr [\hat \partial_y^b \dot U_{x,t} U_{x,t}\tran L_a] +  \Tr [\hat \partial_y^b  U_{x,t} \dot U_{x,t}\tran L_a] \right\}
\end{align}
However, given the flow equation Eq.~\eqref{eq:UflowApp}, we obtain (suppressing the time dependence for simplicity):
 \begin{align}
   \dot{\cal J}^{a,b}[V](x,y) =& - \frac{1}{2} \left\{ \Tr [\hat \partial_y^b Z_x U_x U_x\tran L_a]  +  \Tr [\hat \partial_y^b  U_x U_x\tran Z_x\tran L_a]\right\}\\
   =& - \frac{1}{2} \left\{ \Tr [\hat \partial_y^b Z_x L_a]  + 
   \Tr [Z_x \hat\partial_y^b U_x U_x\tran L_a] + 
   \Tr [\hat \partial_y^b U_x U_x\tran Z_x\tran L_a]\right\}
\end{align}
The generator is antisymmetric, and in addition, given the transformation property of the one-forms, 
\begin{align}
    \hat \partial_x^a  = \sum_{y,b} {\cal J }^{b,a}(y,x) \partial_y^b\,.
\end{align}
As a result
\begin{align}
   \dot{\cal J}^{a,b}[V](x,y) =& - \frac{1}{2} \sum_{c,z}  {\cal J }^{c,b}(z,y) \Tr\left\{ \partial_z^c Z_x L_a  + Z_x \partial_z^c U_x U_x\tran L_a -  \partial_z^c U_x U_x\tran Z_x L_a\right\} \nonumber\\
  =&  -\frac{1}{2} \sum_{c,z}  {\cal J }^{c,b}(z,y) \Tr\left\{ \partial_z^c Z_x L_a  + Z_x L_c \delta_{x,z}  L_a -  L_c \delta_{x,z}  Z_x L_a\right\}\nonumber \\
 =& - \frac{1}{2}\sum_{c,z} {\cal J }^{c,b}(z,y)\Tr\left\{ \partial_z^c Z_x L_a  +\sum_d Z_x\delta_{x,z} \epsilon_{cad} L_d\right\}
\end{align}
where
 \begin{align}
     \partial_z^c U_x U_x\tran= L_c \delta_{xz} 
 \end{align}
  and the $\mathfrak{so}(3)$ algebra of Eq.~\eqref{eq:so3LieAlgebra} was used.
  Putting everything together, we obtain
 \begin{align}
     \Tr [ \dot {\cal J}[U_t] {\cal J}_t^{-1}[U_t]] =& \sum_{x,a}\partial_x^a  Z^a_x 
\end{align}
and if the flow generator is the gradient of a flow action
\begin{align}
    Z^a_x  = {\partial_x^a} S_f[U]  \nonumber
\end{align}
we get
\begin{align}
     \frac{d \ln\det [{\cal J} ]}{dt} =&\; \sum_{x,a}[{\partial_x^a}]^2 S_f[U]
     \label{eq:jacobianODE_app}
 \end{align}

\bibliography{biblio, new_biblio}

\end{document}